\documentclass{aa}  
\usepackage{graphicx}
\usepackage{xcolor}

\usepackage{txfonts}
\usepackage[colorlinks=true, citecolor=blue, linkcolor=violet]{hyperref}
\usepackage{subfig}
\usepackage{pdflscape}

\newcommand{\kms }{km~s$^{-1}$}
\newcommand{\ergs }{erg~s$^{-1}$}

\newcommand{ \Msun } {$M_{\odot}$}

\newcommand{ \Msunyr } {M$_{\odot}$~yr$^{-1}$}

\newcommand{ \Htwo } {H$_2$}
\newcommand{ \Ha} {H$\alpha$}
\newcommand{ \Hb} {H$\beta$}

\newcommand{ \Pab} {Pa$\beta$}
\newcommand{ \Pag} {Pa$\gamma$}
\newcommand{ \Brg} {Br$\gamma$}
\newcommand{ \OIII} {[O\,\textsc{iii}]}
\newcommand{ \NII} {[N\,\textsc{ii}]}
\newcommand{ \SII} {[S\,\textsc{ii}]}

\newcommand{ \OI} {[O\,\textsc{i}]}
\newcommand{ \SIII} {[S\,\textsc{iii}]}
\newcommand{ \FeII} {[Fe\,\textsc{ii}]}
\newcommand{ \PII} {[P\,\textsc{ii}]}
\newcommand{ \HeI} {He\,\textsc{i}}
\newcommand{ \FeVII} {[Fe\,\textsc{vii}]}
\newcommand{ \NeV} {[Ne\,\textsc{v}]}

\newcommand{ \emcee}{{\tt emcee}}
\newcommand{ \gaia}{\textit{Gaia}}
\newcommand{ \Chandra}{\textit{Chandra}}

\newcommand{ \micron}{$\mu$m}

\newcommand{ \Mrate}{$\dot{M}_\mathrm{out}$}

\newcommand{\MCG}{MCG-03-34-64}

\newcommand{\FeKa}{Fe-K$\alpha$}

\begin{document}

   \title{Fast, dust-poor outflows  in the local candidate dual AGN MCG-03-34-64 observed with VLT/ERIS}
   
   \titlerunning{Fast, dust-poor outflows in MCG-03-34-64}


   \author{I.~Lamperti\inst{\ref{iUNIFI},\ref{iOAA}}\fnmsep\thanks{E-mail:    isabella.lamperti@unifi.it}
          \and F.~Mannucci\inst{\ref{iOAA}}
          \and E.~Bertola\inst{\ref{iOAA}}
          \and A.~Marconi\inst{\ref{iUNIFI},\ref{iOAA}}
          \and G.~Cresci\inst{\ref{iOAA}}
          \and E.~Nardini\inst{\ref{iOAA}}
          \and Q.~D'Amato\inst{\ref{iOAA}}
          \and M.~Perna\inst{\ref{iCAB}}
          \and A.~Rojas-Lilayú\inst{\ref{iUTFSM}}
          \and C.~Bracci\inst{\ref{iUNIFI},\ref{iOAA}}
          \and V.~Braito\inst{\ref{iOABrera}}
          \and E.~Cataldi\inst{\ref{iUNIFI},\ref{iOAA}}
          \and M.~Ceci\inst{\ref{iUNIFI},\ref{iOAA}}
          \and A.~Chakraborty\inst{\ref{iOAA}}
          \and C.~Cicone\inst{\ref{iOslo}}
          \and A.~De Rosa\inst{\ref{iIAPS-Roma}}
          \and A.~Feltre\inst{\ref{iOAA}}
          \and M.~Ginolfi\inst{\ref{iUNIFI},\ref{iOAA}}
          \and E.~Lusso\inst{\ref{iUNIFI},\ref{iOAA}}
          \and C.~Marconcini\inst{\ref{iUNIFI},\ref{iOAA}}
          \and B.~Moreschini\inst{\ref{iUNIFI},\ref{iOAA}}
          \and E.~Portaluri\inst{\ref{iOAAb}}
          \and K.~Rubinur\inst{\ref{iOslo}}
          \and M.~Scialpi\inst{\ref{iUNIFI},\ref{iOAA}, \ref{iUniTrento}}
          \and P.~Severgnini\inst{\ref{iOABrera_MI}}
          \and G.~Tozzi\inst{\ref{iMPE}}
          \and A.~Trindade Falc\~{a}o\inst{\ref{iNASA_Goddard}}
          \and L.~Ulivi\inst{\ref{iCAB}}
          \and G.~Venturi\inst{\ref{iSNS}}
          \and C.~Vignali\inst{\ref{iUniBo}, \ref{iOABo}}
          \and M.V.~Zanchettin\inst{\ref{iOAA}}
          }


   \institute{
            Università di Firenze, Dipartimento di Fisica e Astronomia, via G. Sansone 1, I-50019 Sesto F.no, Firenze, Italy\label{iUNIFI}
    \and 
        INAF - Osservatorio Astrofisico di Arcetri, Largo E. Fermi 5, I-50125 Firenze, Italy\label{iOAA}
    \and
        Centro de Astrobiolog\'ia (CAB),
            CSIC--INTA, Cra. de Ajalvir Km.~4, 
            28850 -- Torrej\'on de Ardoz, Madrid, Spain\label{iCAB}
    \and 
        Departamento de F\'isica, Universidad T\'ecnica Federico Santa Mar\'ia, Vicu\~{n}a Mackenna 3939, San Joaqu\'in, Santiago, Chile\label{iUTFSM}
    \and
        INAF - Osservatorio Astronomico di Brera, Via Bianchi 46, I-23807 Merate (LC), Italy\label{iOABrera}
    \and
        Institute of Theoretical Astrophysics, University of Oslo, Postboks 1029, NO-0315 Oslo, Norway\label{iOslo}
    \and
        INAF - Istituto di Astrofisica e Planetologia Spaziali, Via Fosso del Cavaliere 100, 00133 Rome, Italy\label{iIAPS-Roma}
    \and 
        INAF-Osservatorio Astronomico d'Abruzzo,
        Via Mentore Maggini, I-64100 Teramo, Italy\label{iOAAb}
    \and
        University of Trento, Via Sommarive 14, I-38123 Trento, Italy\label{iUniTrento}
    \and
        INAF - Osservatorio Astronomico di Brera, via Brera 28, 20121
        Milano, Italy\label{iOABrera_MI}  
    \and 
        Max Planck Institute for extraterrestrial Physics, Giessenbachstraße 1, D-85748 Garching, Germany \label{iMPE}
    \and 
        NASA Goddard Space Flight Center, Code 662, Greenbelt, MD 20771, USA
        \label{iNASA_Goddard}
    \and 
        Scuola Normale Superiore, Piazza dei Cavalieri 7, I-56126 Pisa, Italy\label{iSNS}
    \and
        Dipartimento di Fisica e Astronomia, Alma Mater Studiorum, Università degli Studi di Bologna, Via Gobetti 93/2, I-40129 Bologna, Italy\label{iUniBo}
    \and
        INAF – Osservatorio di Astrofisica e Scienza dello Spazio di Bologna, via Gobetti 93/3, I-40129 Bologna, Italy\label{iOABo}
        }

   \date{Received 27 January 2026 / Accepted 19 March 2026}

 
  \abstract
   {
    
    We present VLT/ERIS integral field unit (IFU) J-band observations of \MCG,  a nearby ($z= 0.0167$) Seyfert galaxy hosting a candidate dual AGN system with a separation of $\sim100$~pc between the nuclei, which are detected in \FeKa\ X-ray images and \OIII$\lambda$5007\AA\ narrow-band images. 
    The ERIS observations cover, among others, the \HeI$\lambda$1.083\micron, \FeII$\lambda$1.257\micron\ and \Pab\ emission lines, over a field of view of $3''\times3''$ ($\sim1\times1$~kpc$^2$). 
    We analyse the ionised gas kinematics and identify two regions with enhanced velocity dispersion ($W80\sim 1500$~\kms), suggestive of fast outflowing gas, spatially coincident with the position of the two candidate active nuclei. 
    We extract the spectra from the two regions, and use them to derive the properties of the ionised outflows. The two spectra show a prominent blueshifted wing with velocities  $v_{max}\sim-1700$~\kms, corresponding to the highest $2-5$ percentiles of samples of local AGN with similar bolometric luminosities ($\sim10^{44}$~\ergs).
     For the ionised phase of the two outflows, we derive comparable  masses of $\sim(4\pm1)\times 10^5$~\Msun\ and mass outflow rates of $\sim20\pm5$~\Msunyr.
    The two distinct outflows could be associated with the two nuclei, or be generated by the interaction of the radio jet with the ISM.
    The map of the \FeII/\Pab\ line ratio  shows the highest values in the more external regions 
    (compatible with shock excitation), intermediate values in the vicinity of the nuclei (compatible with AGN excitation), and lower values between the two nuclei, close to the peak of the optical and NIR continuum. 
    We interpret this as  due to a contribution to the \Pab\ emission of a nuclear starburst  or to enhanced AGN ionisation.
    We also analyse the peculiar (very broad, boxy) profile of the \FeVII$\lambda$6087\AA\ optical coronal line from an archival VLT/X-shooter spectrum.  
    The comparison with the \NeV$\lambda$3425\AA\ and \FeII$\lambda$1.257\micron\ profiles indicates that \FeVII\ emission likely arises only from the outflow.
    The absence of the systemic component in \FeVII\ -- unlike in \NeV, which has similar ionisation potential and critical density -- suggests suppression of \FeVII\ due to iron depletion onto dust grains, while its detection in the outflow implies a lower dust content than in the host ISM.
   The additional information gained from the ERIS data are consistent with the scenario of a dual AGN,  however further observations are required to confirm its nature.
    }


   \maketitle
%
\section{Introduction}


Supermassive black holes (SMBHs) are believed to occupy the centre of most massive galaxies \citep[e.g.,][]{Magorrian1998}. Therefore, the merger of two massive galaxies is expected to result in a system containing two SMBHs \citep[e.g.][]{Koss2012, Koss2023}. 
Hydrodynamical simulations suggest that major mergers can drive gas inflows toward galactic centres, triggering both star formation and accretion onto SMBHs \citep{Mayer2007}. 
This accretion can cause the two SMBHs to shine as active galactic nuclei (AGN) \citep[e.g.][]{Koss2012}.
Active SMBH pairs with small separation (typically $<10$~kpc) are called `dual AGN'; `binary AGN', on the other hand, are those so close that they are within their mutual gravitational sphere of influence \citep[separation $\lesssim100$~pc, depending on the masses,][]{Burke-Spolaor2014}.

The presence of dual AGN within galaxies is a prediction of current models of galaxy and black hole co-evolution \citep[e.g.][]{Tremmel2017, Volonteri2022} and studying these systems is therefore an important observational test.
Moreover, binary AGN are expected to eventually merge into a single SMBH, generating gravitational wave signals that can be detected by pulsar timing arrays (PTA) \citep[e.g.][]{Arzoumanian2018,  Agazie2023a, EPTACollaboration2023} and the future Laser Interferometric Space Antenna \citep[LISA, e.g.][]{Amaro-Seoane2023, Colpi2024arxiv}. 
A detailed characterisation of dual AGN properties and demographics is thus needed to better understand the initial conditions of the SMBHs merging process, constrain the formation path of SMBHs in the core of massive galaxies \citep[e.g.,][]{Sicilia2022,Amaro-Seoane2023}, and refine expectations about the gravitational waves event rate \citep[e.g.,][]{DeGraf2024,Dong-Paez2023, Chen2023}.

Many dual AGN have been identified serendipitously \citep[e.g.,][]{Junkkarinen2001, Bianchi2008_Mrk463, Koss2011b, Huang2014, Husemann2018a, Balmaverde2018,  Matsuoka2024, Perna2025, Zamora2025, Merida2025}. 
Others have been discovered through systematic searches using a variety of techniques:
for example, looking for double peaks in the emission line profiles \citep{Zhou2004,  Comerford2009};
detecting multiple peaks in the \textit{Gaia} satellite's 1D light profiles \citep{Mannucci2022, Mannucci2023}; observing astrometric shifts in the photo-centres of quasars observed by \textit{Gaia} \citep{Shen2019a, Hwang2020, Chen2022}; looking for multiple \textit{Gaia} sources close to known AGN \citep{Lemon2017,Lemon2023}; identifying multiple bright nuclei in the optical, near-infrared or millimetre continuum images of merging galaxies \citep[e.g.][]{Liu2011, Koss2012, Koss2018, Koss2023}; or detecting close companions to AGN in high-resolution Euclid images \citep{Ulivi2025arXiv}.
 In general, all these techniques require follow-up observations with X-ray, radio, or resolved optical/NIR spectroscopy to confirm the dual AGN nature of the systems \citep[e.g.][]{Comerford2011, McGurk2011, Mueller-Sanchez2015, Satyapal2017, Rubinur2019, DeRosa2023, Ciurlo2023, Chen2023_dual, Scialpi2024, Chen2024_dual, Ishikawa2025}.

Identifying small-separation 
($\delta< 1$~kpc)
dual AGN is challenging due to the high angular resolution required to resolve the two nuclei and the possible increased obscuration in late-stage mergers \citep[e.g.][]{Koss2016b, Ricci2021, DeRosa2023}.
Until now, only a handful of dual or binary SMBHs have been confirmed at $\delta< 1$~kpc (e.g.  4C+37.11, \citealt{Rodriguez2006}; NGC 6240, \citealt{Komossa2003, Fabbiano2020, Kollatschny2020, Treister2020}; NGC 7727,  \citealt{Voggel2022}; UGC 4211, \citealt{Koss2023}), 
while a few other candidates have been proposed (e.g. \citealt{Wang2009}, \citealt{Woo2014}; \citealt{Mueller-Sanchez2015}; \citealt{Graham2015}, but see also \citealt{Rigamonti2025}; \citealt{Kharb2017}, but see also \citealt{Breiding2022}; \citealt{Severgnini2018}, 
\citealt{Serafinelli2020},  but see also \citealt{Foord2025}; \citealt{Severgnini2021}; \citealt{Fatovic2025},  but see also \citealt{Rigamonti2025b}).
In any case, multi-wavelength observations are crucial to confirm the presence of dual AGN systems, as demonstrated by studies such as, e.g., \citet{DeRosa2019} and  \citet{Rubinur2021}.

AGN can produce strong gas outflows, detected in different phases (molecular, ionised, neutral) and on various spatial scales \citep[e.g.][]{Cicone2018}. These outflows can affect the star formation in the host galaxy, by removing, heating or compressing the gas, a process which is usually referred to as AGN feedback \citep[e.g.][]{ Harrison2017, Cresci2018_review, Harrison2024_review}.
Simulations and observations of galaxy mergers show that the dual AGN phase coincides with a critical phase in which both SMBH accretion and star formation activity are more powerful \citep[e.g.][]{VanWassenhove2012, Koss2012, Blecha2013}.
In fact, outflows have been detected in several dual AGN up to $z\sim3$ \citep[e.g.][]{Cicone2018b, Saturni2021, Tubin2021, Ceci2025, Carlsen2025arXiv, Hermosa-Munoz2025, Bertola2025}.





Recently, \citet{TrindadeFalcao2024} discovered a candidate dual AGN in the nearby ($z=0.017$) galaxy \MCG, combining \OIII$\lambda$5007, X-ray and radio observations.
If confirmed, this would be one of the closest known dual AGN, with a separation between the two nuclei of $\sim 100$~pc.
In this work, we analyse new observations of \MCG\ obtained with SPIFFIER, the near-infrared (NIR) integral field unit (IFU) of the Enhanced Resolution Imager and Spectrograph \citep[ERIS,][]{Davies2023} at the Very Large Telescope (VLT).
These data allow us to study the ionised gas kinematics in the nuclear region of this galaxy and investigate the presence of outflows, possibly related with the nuclei.

The paper is organised as follows.
Section~\ref{sec:obs_datared} presents the target, the VLT/ERIS observations, and the ancillary data from the Hubble Space Telescope (HST), VLT/X-shooter, the Very Large Array (VLA), and the {\it Chandra} X-ray Observatory.
In Sec.~\ref{sec:analysis} we describe the method used for the emission line fitting.
In Sec.~\ref{sec:results} we present the results: we show the ionised gas emission maps and kinematics (Sec.~\ref{sec:emission_line_maps}), the properties of the outflows (Sec.~~\ref{sec:outflows}) and their comparison with the literature (Sec.~\ref{sec:outflow_literature}). In Sec.~\ref{sec:line_ratios} we show emission line ratio diagnostics used to identify the main excitation mechanisms. Section~\ref{sec:radio} and Sec.~\ref{sec:Xray_spectra} present the radio continuum data and the X-ray spectra, respectively. In Sec.~\ref{sec:coronal_lines} we discuss the peculiar profile of the \FeVII\ optical emission line.  
Finally, in Sec.~\ref{sec:discussion} we discuss the evidence for a dual or single AGN scenario, and  in Sec.~\ref{sec:conclusions} we summarize our conclusions.

Throughout this work, we assume a cosmological model with $\Omega_{\Lambda} = 0.7$, $\Omega_{\text{M}}= 0.3$, and $H_0 = 70$ km s$^{-1}$ Mpc$^{-1}$, corresponding to an angular scale of 340 pc arcsec$^{-1}$ at the redshift of MCG-03-34-64.







\section{Target description, observations and data reduction}
\label{sec:obs_datared}

\subsection{Target description} 
\MCG\ (R.A. 13h22m24.4618s, Dec.$-16^{\circ}$43\arcmin42.497\arcsec), 
also known as IRAS 13197-1627, is a luminous infrared galaxy \citep[$L_{8-1000~\mu\rm m} = 1.7\times 10^{11}~L_{\odot}$,][]{Sanders2003} at $z = 0.01672$. 
It has been detected by the Burst Alert
Telescope (BAT) instrument on board of the Swift telescope \citep{Baumgartner2013}, and is part of the BAT AGN Spectroscopic Survey (BASS) sample \citep{Koss2017, Koss2022}.
\MCG\ has an intrinsic X-ray luminosity of $\log L_{2-10\text{ keV}}/[\text{erg s}^{-1}] = 43.41$ and is highly obscured, with a column density of $\log N_{\rm H}/[\text{cm}^{-2}] = 23.8\pm0.02$ \citep{Ricci2017}.  It was originally classified as a Seyfert~1.8 \citep{Aguero1994, Oh2022}, 
indicating the presence of a broad-line region (BLR) component in the \Ha\ line and a weak BLR component in \Hb\ (but see Sec.~\ref{sec:xshooter_fit}). \MCG\ has a radio jet \citep{Schmitt2001, Song2022}, and shows a radio excess relative to the radio-FIR correlation \citep{Condon1995, Corbett2003}.

\begin{figure*}[!t]
\centering 
\includegraphics[width=0.6\textwidth, trim=0mm 5mm 0mm 0mm, clip]{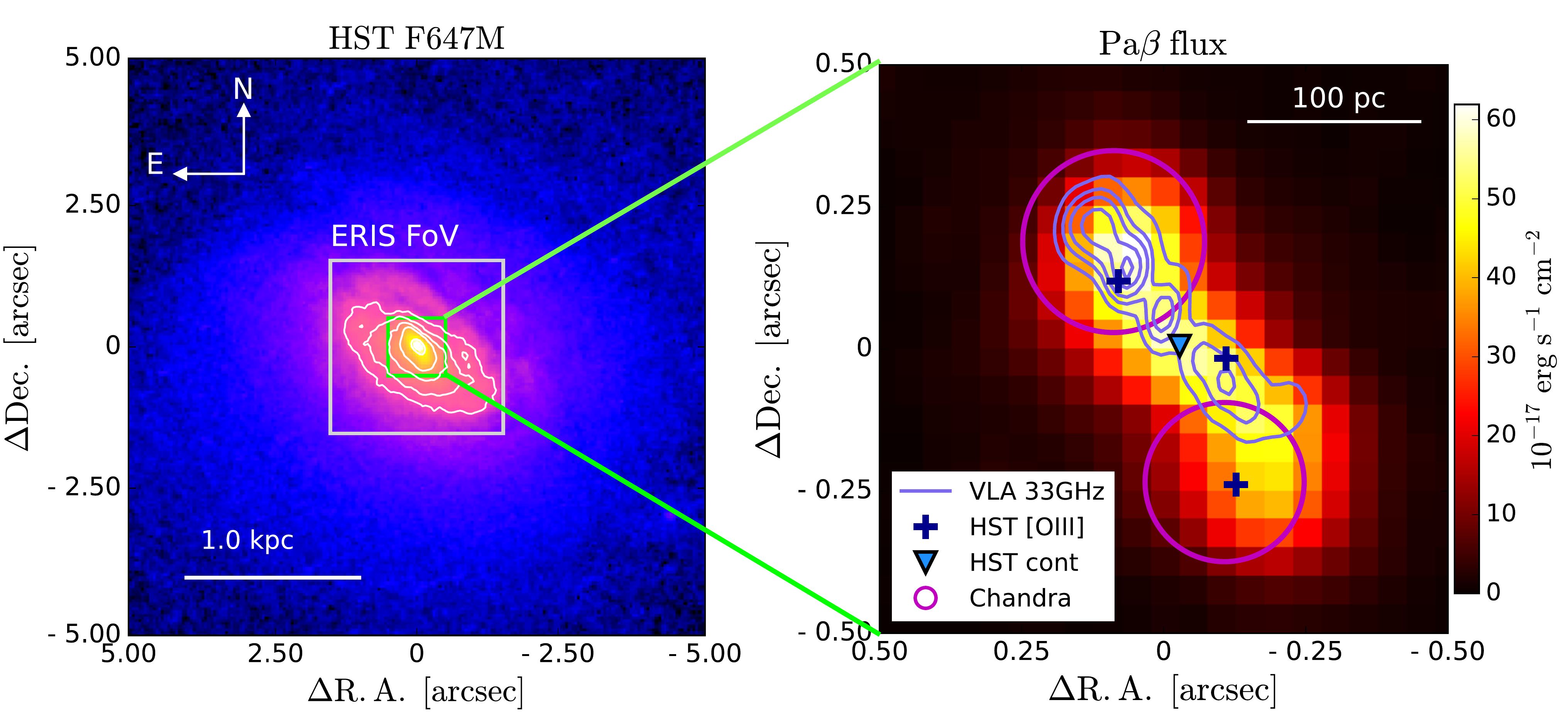}
\includegraphics[width=0.32\textwidth, trim=0mm 3mm 0mm 0mm, clip]{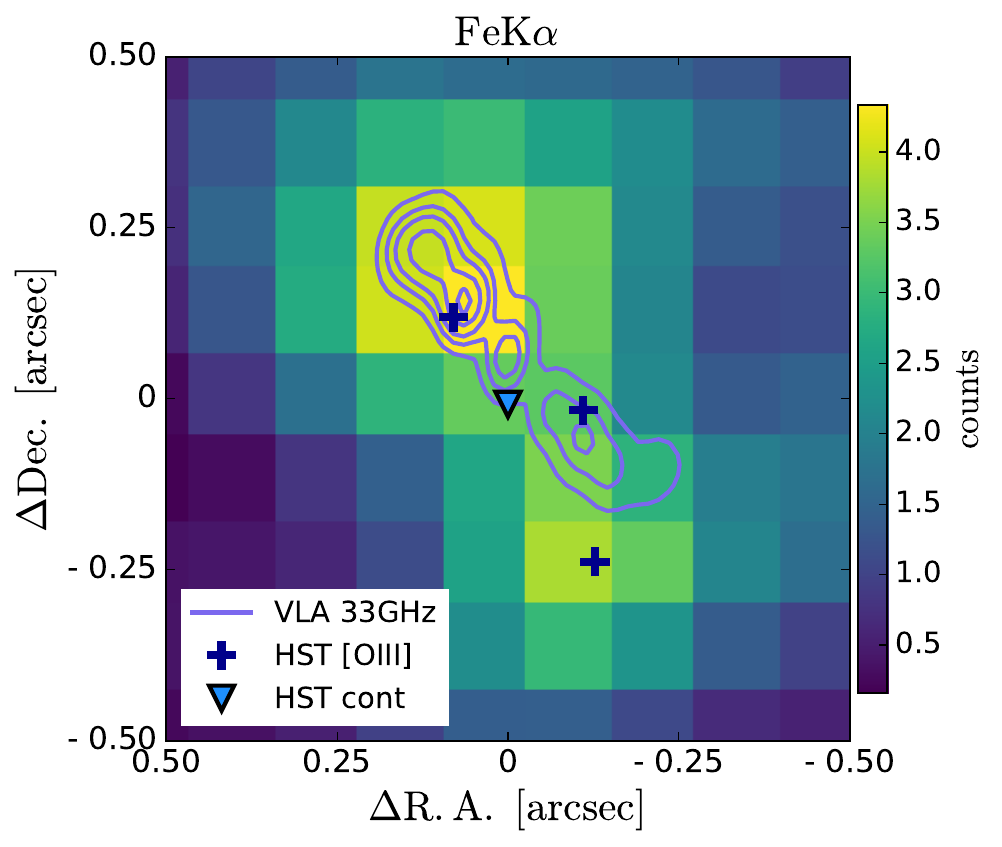}
\caption{HST, ERIS and \Chandra\ images of \MCG. \textit{Left:} HST F647M continuum image, with contours at arbitrary flux levels. 
The grey rectangle shows the $3''\times3''$ ERIS FoV, while the green rectangle shows the size of the middle and right panels ($1''\times1''$). North is up and east to the left.
\textit{Middle:} Map of the total \Pab\ flux from ERIS, showing the position of the sources detected at different wavelengths: the \OIII\ peaks from HST are shown with blue crosses (+), 
the X-ray {\it Chandra} peaks with magenta circles indicating the position uncertainties, 
 the HST continuum centroid (and ERIS continuum peak) with a lightblue triangle. 
Lavender contours show the radio emission at 33 GHz from VLA. \textit{Right:} Map of the \FeKa\ emission from \Chandra. The image refers to the 6.2--6.6 keV rest-frame band at the 1/4 subpixel level, with a Gaussian smoothing with standard deviation of 1 image pixel.} 
\label{fig:nuclei_position}
\end{figure*}

\citet{TrindadeFalcao2024}  identified three emitting peaks in the HST \OIII\ narrow-band image. The analysis of {\it Chandra} X-ray data further revealed two peaks in the $6.2-6.6$~keV energy band, which traces the \FeKa\ fluorescence line, suggesting the presence of more than one AGN (see Fig.~\ref{fig:nuclei_position}).
Archival VLA radio data also show two peaks of emission spatially coincident with the X-ray peaks, although these are unresolved. The authors propose two possible interpretations. 
In the first scenario, a single AGN is located at the position of the northern \OIII, \FeKa, and radio peaks, while the other two \OIII\ peaks are attributed to interactions between the interstellar medium (ISM) and a radio jet \citep{Schmitt2001}. Previous X-ray studies on this target report evidence for a mixture of collisionally ionised and photo-ionised gas in the narrow-line region, which could support this scenario \citep{Miniutti2007}.
However, the similarly high \FeKa\ luminosities measured in the individual regions led the authors to disfavour this scenario.
The second possibility is the presence of a dual AGN system, with a projected separation $125\pm21$~pc (0.37\arcsec), together with shocked ISM that would account for the third \OIII\ peak. This scenario is supported by the tentative detection of two spatially resolved \FeKa\ emitting regions, with comparably high line luminosities ($L_{\rm FeK\alpha}> 10^{40}$~erg~s$^{-1}$), 
as well as by the presence of spatially coincident \OIII\ and radio centroids.

Based on the \OIII\ luminosity, \citet{TrindadeFalcao2024} inferred bolometric luminosities of $1.6\times 10^{44}$~\ergs, $7.3\times 10^{43}$~\ergs\ and $1.0\times 10^{44}$~\ergs\ for the northern, central and southern \OIII\ peaks, respectively, adopting a bolometric correction factor $c = 454$ \citep{Lamastra2009}.

\subsection{ERIS observations}
\MCG\ was observed with
 ERIS \citep{Davies2023} on ESO/VLT as part of the Guaranteed Time Observations (GTO) program P113.26BW.001 (P.I.~F.~Mannucci).
In particular, we employ the SPIFFIER IFU spectrograph \citep{George2016}, which is an upgrade of the SPIFFI spectrograph of SINFONI \citep{Eisenhauer2003}. We use the J$\_$low grating, covering the wavelength range $1.09-1.42$~\micron\ with a resolving power R$\sim5000$  (corresponding to $\sigma \sim25$~\kms). 
The target was observed on  April 5th 2024 in adaptive optic (AO) mode with the Laser Guide Star (LGS) and using the galaxy nucleus itself as tip-tilt star.
We used a spaxels size of 100$\times$50~mas with half-pixel dithering. 
The observations were taken in a single observing block, with eight exposures on target and four sky frames of 120~s each, for a total exposure time of 24~min (16~min on source). The individual exposures were dithered including shifts of half a spaxel, enabling the recovery of an effective spaxel size of 50~mas during data reduction.

Data reduction was carried out with the ERIS pipeline v1.8.1 using EsoRex \citep{EsoRex2015}. 
Single frames were corrected for differential atmospheric refraction, sky subtraction was performed following the method of \citet{davies2007_skysub}, as implemented in the ERIS pipeline, and setting {\sf stretch=True}. We produce the mean coadded cube by combining the single frames with the SINFONI-like combine recipe available in the ERIS pipeline ({\sf eris\_ifu\_combine}) clipping outliers at 2$\sigma$. 
Flux calibration and telluric correction were applied using the dedicated observation of a telluric star (HIP68372), observed right after \MCG. The response function was obtained from the spectrum of HIP68372 normalised to its J magnitude, taken from the 2MASS catalogue. 
The J\_low band is included within the 2MASS J band, allowing  full recovery of the flux spectral distribution.
The final datacube has a spaxel size of 50~mas ($\sim17$~pc) and a field of view of $3.1\times3.1$~arcsec$^2$, corresponding to approximately $1\times1$~kpc$^2$ at the redshift of the target.

 Since no point source is present in the field, and no PSF star was observed together with our target, we estimate an upper limit on the PSF FWHM of $\sim0.15\arcsec$ from the size of the most compact structure detected in the field-of-view, i.e. the blueshifted outflow in the north, see Sec.~\ref{sec:outflows}.

\subsection{Ancillary data}
\label{sec:ancillary_data}


\paragraph{HST/ACS$\_$WFC}images for \MCG\ are presented in \citet{TrindadeFalcao2024}. We downloaded the HST data available for this target from the Barbara A. Mikulski Archive for Space Telescopes (MAST)\footnote{\href{https://mast.stsci.edu/portal/Mashup/Clients/Mast/Portal.html}{https://mast.stsci.edu/portal/Mashup/Clients/Mast/Portal.html}}.
Observations have been obtained in June 2022 (proposal ID: 16847, P.I.: T.~J.~Turner) 
with the narrow filter FR505N centred at 5090\AA \ observed-frame (bandwidth 100\AA\ or 2\%) targeting the \OIII\ emission line;  and with the medium filter FR647M, centred at 5595~\AA\ observed-frame 
(bandwidth 505~\AA\ or 9\%), targeting the continuum. 
The left panel of Fig.~\ref{fig:nuclei_position} presents the HST FR647M optical continuum image of \MCG, with the ERIS field of view overlaid.
 

\vspace{-\baselineskip}
\paragraph{VLT/X-shooter}observations of \MCG\ were obtained on March 25th 2019 (program 0102.A-0433, P.I: B.~Trakhtenbrot), with an exposure time of 480~s in seeing-limited mode (seeing full width at half maximum (FWHM) $1.1-1.4$\arcsec). 
The observations include the three X-shooter arms UVB, VIS and NIR, which have slit sizes of $1.6''\times 11''$,  $1.5''\times 11''$, and $0.9''\times 11''$, respectively. The resolving powers of the three bands are R = 3200, 5050 and 5000, respectively\footnote{\href{https://www.eso.org/sci/facilities/paranal/instruments/xshooter/inst.html}{www.eso.org/sci/facilities/paranal/instruments/xshooter/inst.html}}. 
The slits are oriented with a position angle of $-18$~deg (east-of-north), which cover the positions of the three \OIII\ peaks, but  are unable to resolve them since the distance between the north and south peaks is $\sim0.4''$, smaller than the seeing at the time of the observations.
We downloaded the reduced 1D X-shooter spectra from the ESO archive.

\vspace{-\baselineskip}
\paragraph{VLA} observations of \MCG\ in A configuration, which provides the highest spatial resolution, have been obtained at 8.46, 15 and 33~GHz. The VLA-A 8.46~GHz observations used in \citet{TrindadeFalcao2024}  have a synthetized beam FWHM$\sim0.3\arcsec \times 0.2\arcsec$. 
Here, we analyse the 15 and 33~GHz observations, which have higher spatial resolution.
They have been obtained as part of the program 14A-471 (P.I: A.~Evans) 
and have been presented in \citet{Song2022}.
The Ku band (15~GHz) and Ka band (33~GHz) have  been observed on March 7th and 23rd 2014, respectively.
We downloaded the data from the VLA NRAO archive, and performed the calibration using the standard EVLA pipeline version 1.4.2, with the Common Astronomy Software Applications ({\tt CASA}) v5.3.0, \citep[][]{CASA}.
The data were cleaned using the \textsc{tclean} CASA task with a \textsc{Briggs} weighting scheme with \textsc{robust} = 0.5.
This results in a  beam FWHM of $0.24\arcsec \times 0.11\arcsec$ (position angle P.A.= 30$^{\circ}$ east-of-north) for the 15~GHz image, 
and  $0.09\arcsec \times 0.05\arcsec$ (P.A.= 167$^{\circ}$) for the  33~GHz image.
The final reduced images have a root mean square (RMS) of 88~$\mu$Jy beam$^{-1}$ and 24~$\mu$Jy beam$^{-1}$, respectively. The astrometric precision is approximately FWHM$_\text{beam}$/(2$\times$SNR) \citep{Condon1997}, which translates into uncertainties < 0.01\arcsec\ for these observations. The radio images are shown in Fig.~\ref{fig:radio_images}.


\vspace{-\baselineskip}
\paragraph{\textit{Chandra}}
Observations of \MCG\ were performed in 2023 with the Advanced CCD Imaging Spectrometer (ACIS) onboard {\it Chandra}, and were already presented by \citet{TrindadeFalcao2024}. The observations were split over three consecutive days: April 19th (ObsID 25253; exposure 15.59 ks), 20th (ObsID 27802; exposure 17.85 ks), and 21st (ObsID 27803; exposure 16.89 ks). 
We retrieved the event files from the High Energy Astrophysics Science Archive Research Center (HEASARC)\footnote{\href{https://heasarc.gsfc.nasa.gov/}{https://heasarc.gsfc.nasa.gov/}}, and reprocessed them with the {\it chandra\_repro} script within the \texttt{CIAO} 4.17 software package\footnote{\href{https://cxc.cfa.harvard.edu/ciao/}{https://cxc.cfa.harvard.edu/ciao/}}, using the version 4.12.2 of the Calibration Database (CALDB).

For targets along the optical axis, the size of the ACIS pixel (0.492\arcsec) is larger than the core of the point spread function (PSF) of the High Resolution Mirror Assembly, hence to explore physical scales as small as the separation between the two possible AGN we need to resort to a subpixel image analysis, made it possible by the energy-dependent subpixel event repositioning (EDSER) algorithm \citep{Li2004} that is implemented in the default reduction pipeline. We then had to correct for the relative astrometry of the three {\it Chandra} exposures before merging them into a single image. Whenever possible, the astrometric correction is performed by considering the position of all the point-like sources in the field of view. Yet, as already noted by \citet{TrindadeFalcao2024}, the only other two detections around MCG-03-34-64 are too faint and located more than 1\arcmin\ away near the edge of the chip array to be reliable. We therefore used the source itself, computing for each image the centroid over the 0.5--7 keV band within 10 pixels (4.92\arcsec) from the visual peak.

We took ObsID 27802 as a reference, as this is marginally longer and proportionally contains more counts. The required shifts $[\Delta x,\Delta y]$ in sky coordinates relative to ObsID 27802 were $[-0.8,-1.0]$ and $[+0.9,+0.5]$ pixels for ObsID 25253 and ObsID 27803, respectively. These shifts are consistent within $[\pm0.1,\pm0.2]$ pixels with those obtained by \citet{TrindadeFalcao2024}, who considered the hard-band (2--7 keV) images instead\footnote{For a sanity check, we also computed the source centroids in the 2--7 keV and iron-K (6--7 keV) bands, confirming that the modulus of the uncertainty on the relative shifts between the three images is less than 0.25 pixels.}, and well within the absolute astrometric accuracy of {\it Chandra}. Indeed, for sources observed near the aimpoint in the last five years (from January 6th 2021 to January 6th 2026) and reprocessed with the latest calibration, the radius of the uncertainty circle of {\it Chandra} ACIS coordinates is 0.66\arcsec (0.88\arcsec) at the 68\% (90\%) level of the cumulative offset distribution with respect to optical/radio counterparts\footnote{\href{https://cxc.cfa.harvard.edu/cal/ASPECT/celmon/}{https://cxc.cfa.harvard.edu/cal/ASPECT/celmon/}}.
After correcting the relative astrometry with {\it wcs\_update}, we combined the three event files with {\it merge\_obs} and obtained a single image with an exposure of 50.33 ks. 
The fine structures that can be appreciated at the 1/8 subpixel level in the 6.2--6.6 keV rest-frame band (centred on the 6.4-keV neutral Fe-K line) are in full agreement with those presented in \citet{TrindadeFalcao2024}. In this work, however, we will only employ the 1/4 subpixel image (0.123\arcsec, right panel of Fig.~\ref{fig:nuclei_position}), as this is completely invariant with respect to the exact criterion adopted for the astrometric correction of the original observations.

Although the Fe-K$\alpha$ emission map is clearly elongated in the north-east to south-west direction, and reminiscent of the morphology in the same band of the prototypical, X-ray-confirmed dual AGN NGC 6240 \citep[e.g.,][]{Fabbiano2020}, the data quality is too limited to draw any safe conclusion about the presence of two separate nuclei. In order to gain some deeper insights into the nature of this system, we also attempted to extract a spectrum for each Fe-K$\alpha$ centroid, adopting circular extraction regions with radii of 0.25\arcsec, slightly off-centred from each peak to avoid superposition. The background was evaluated on a nearby region of 5\arcsec-radius devoid of diffuse X-ray emission. The extraction process was performed with the {\it specextract} script for the three individual exposures, and the resulting source and background spectra, and the relative redistribution matrix and auxiliary response files were then combined with the standard HEASARC tools. The results are briefly discussed in Section~\ref{sec:discussion}.

\begin{figure*}[!]
\centering 
\includegraphics[width=0.98\textwidth, trim=0mm 3mm 0mm 0mm, clip]{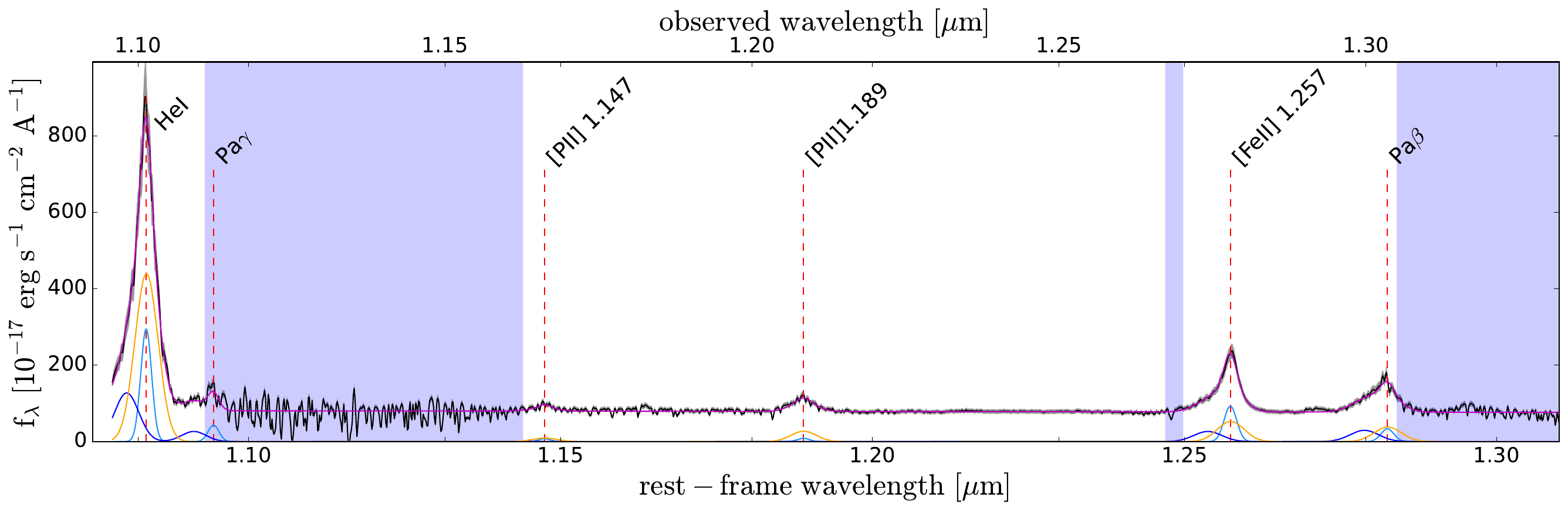}
\caption{Integrated ERIS spectrum of \MCG\ extracted from an aperture of  $r=0.5$\arcsec centred at the continuum peak. The blue-shaded regions show the part of the spectrum affected by telluric absorption. Vertical lines show the wavelength of the main emission lines. The best-fit model is shown in magenta, with individual Gaussian components (narrow, broad1, and broad2) shown in light blue, blue and orange, respectively.}
\label{fig:int_spectrum}
\end{figure*}

\subsection{Astrometry}
\label{sec:astrometry}

We first check that the astrometry of the HST images is aligned with \gaia. We retrieve the positions of the stars in the HST field of view from the \gaia\ catalogue, adjusted to the epoch of the HST observations (2024). Then, we measure the positions of the stars in the HST images using a two-dimensional elliptical Gaussian fit, we compare them with the \gaia\ coordinates and compute the average offset of all the stars.  For the \OIII\ images, we derive an average offset of ($0.03\pm0.02$)\arcsec\ in R.A. and ($0.03\pm0.01$)\arcsec\ in Dec., corresponding to less than one spaxel (spaxel size 0.05\arcsec). For the continuum image, we derive a similar offset of  ($0.02\pm0.02$)\arcsec in R.A. and ($0.02\pm0.01$)\arcsec\ in Dec.

To compare the ERIS maps with the positions of the \OIII\ emission peaks identified in the HST FR505N image by \citet{TrindadeFalcao2024}, we align the ERIS continuum map with the HST FR647M continuum image, assuming that the optical and NIR continuum peak at the same position.
We construct the ERIS continuum map by collapsing the datacube over a wavelength range ($1.24-1.26$~\micron, observed frame) 
that is free from both lines emission and telluric absorption features. 
Then, we align the peak of the ERIS continuum to the peak of the HST continuum image, by applying a shift of 1.39\arcsec\ in R.A. and --1.07\arcsec\ in Dec. to the ERIS astrometry. We estimate the uncertainty to be $\sim1$~spaxel, i.e. 0.05\arcsec. 
The coordinates of the continuum peak are at R.A. 13h22m24.4626s, Dec. --16$^{\circ}$43'42.509\arcsec. The aligned HST and ERIS continuum images are shown in Fig~\ref{fig:HST_ERIS_astrometry} in the Appendix~\ref{sec:appendix_astrometry}.

Given the similar morphology of the radio and \Pab\ emission, we align them by shifting the radio image by $\Delta$Dec=-50mas (consistent with astrometric uncertainties). 
To correct the astrometry of the X-ray image, we align the peak emission in the north to the peak of the 33 GHz map. This corresponds to a shift of 0.08\arcsec\ in R.A. and $-0.25$\arcsec\ in Dec. In the right panel of Fig.~\ref{fig:nuclei_position}, we show the map of the \FeKa\ emission with overlaid the radio contours. The position of the X-ray centroids are indicated in the middle panel, with the circles indicating the uncertainties. These have been estimated based on an enclosed count fraction of 50\% within a distance of 0.25$\pm$0.05\arcsec\ from each peak, and correspond to a circle with a radius of $\approx$\,0.16\arcsec\ for the north source and 0.14\arcsec\ for the south one, respectively. 
 These centroids positions are in full agreement with those reported by \citet{TrindadeFalcao2024}, who aligned the X-ray coordinates to the VLA 8.46~GHz coordinates. The only notable difference is that our  southern X-ray peak is marginally shifted further south by $\sim0.08$~\arcsec\ with respect to their position,  resulting in a better agreement with the location of the southern \OIII\ peak. 

\section{Analysis}
\label{sec:analysis}

 \begin{figure*}[t!]
   \centering
   \begin{minipage}{0.7\textwidth}
   \includegraphics[width=0.5\textwidth, trim=0mm 2mm 0mm 0mm, clip]{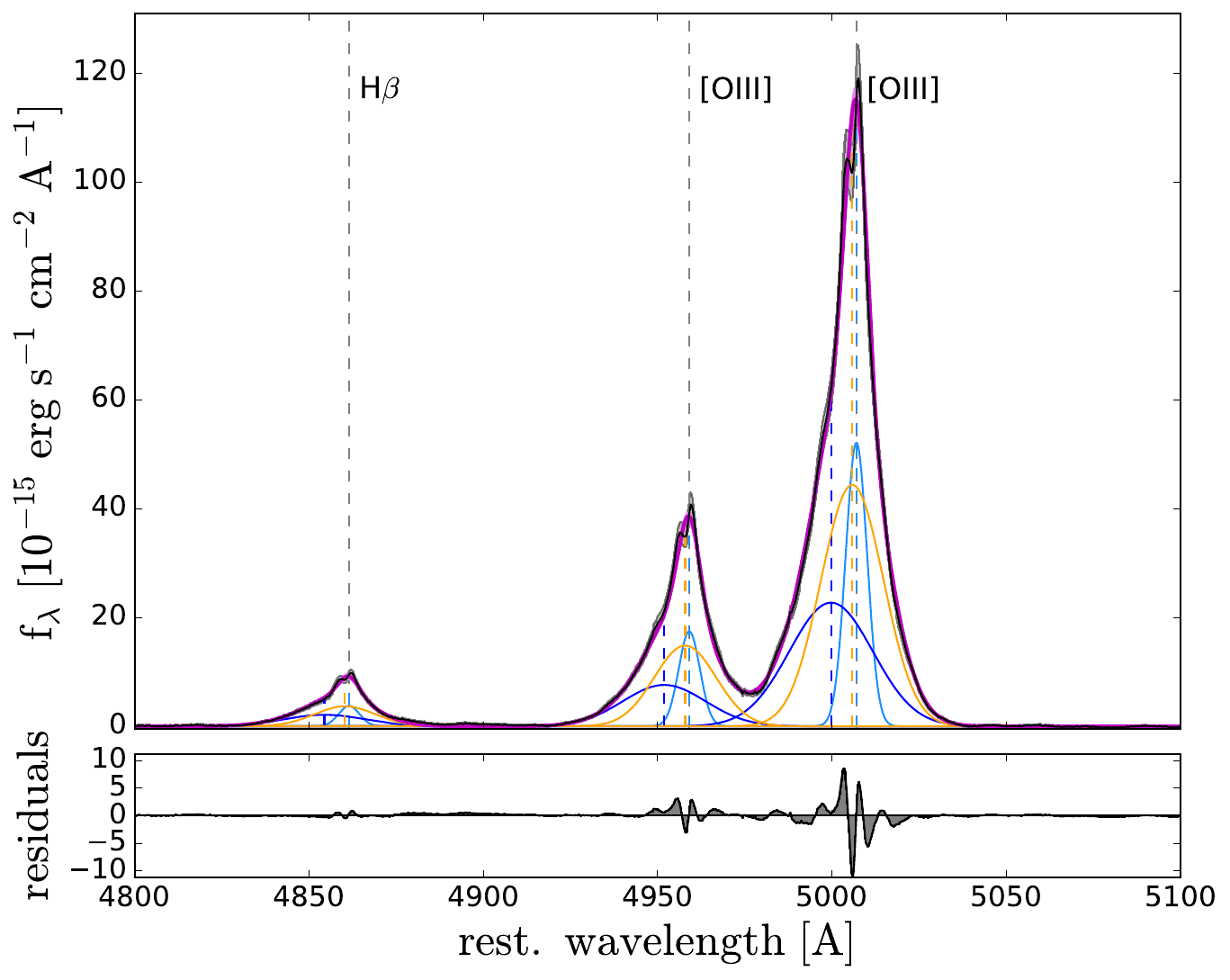}
\includegraphics[width=0.49\textwidth, trim=0mm 2mm 0mm 0mm, clip]{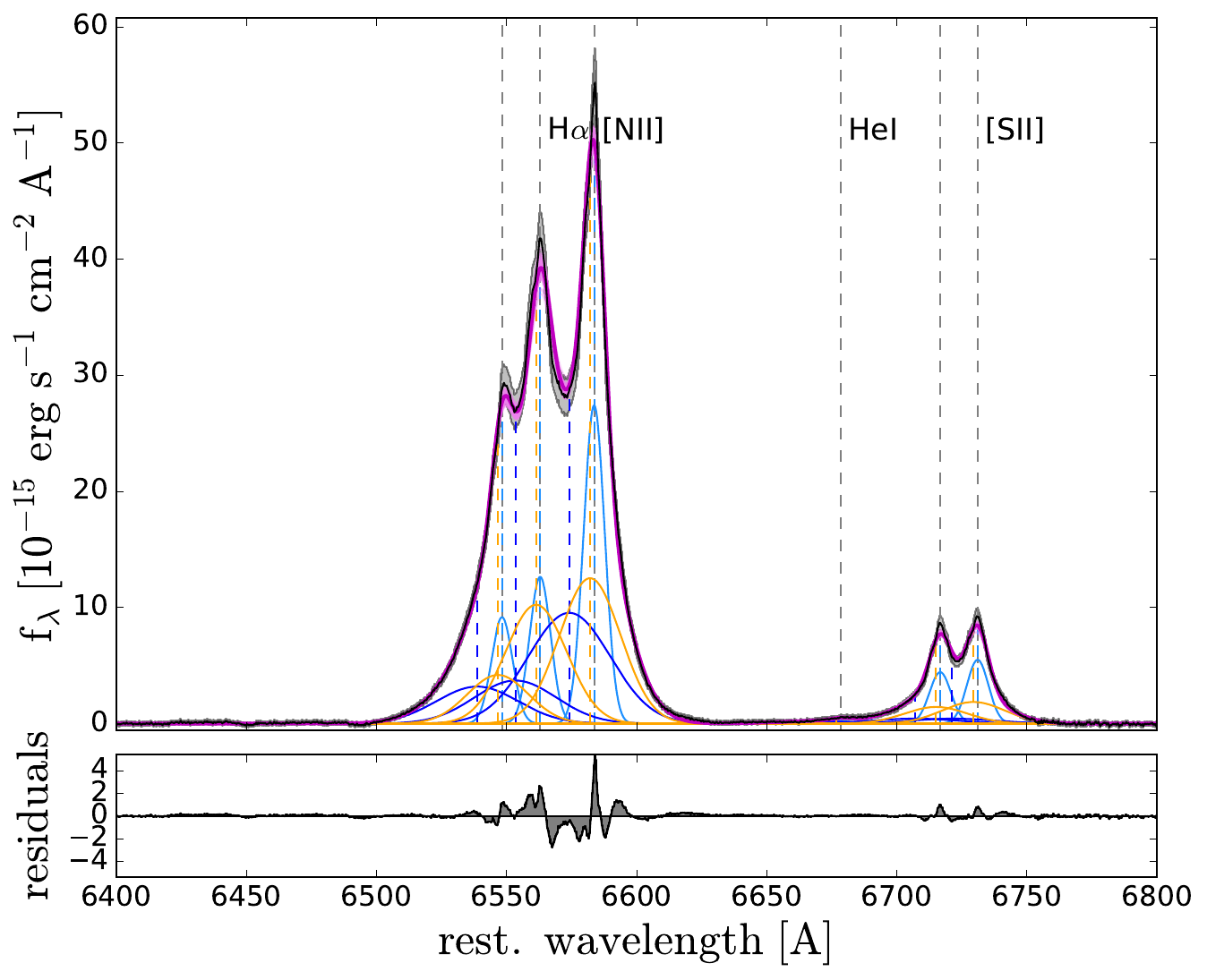}
   \end{minipage}
   \hfill
  \begin{minipage}{0.28\textwidth}
   \caption{Fit of the  optical X-shooter spectrum. \textit{Left:} \Hb\ and \OIII\ lines. \textit{Right:} \Ha, \NII\ and \SII\ lines. Observed spectrum  is in black; the magenta curve shows the best-fit total profile; the light-blue, blue and orange curves show the narrow, b1 and b2 components, respectively. The vertical dashed lines show the centroid positions of the difference components. The bottom panel shows the residuals.}
   \label{fig:spectra_opt}%
    \end{minipage}
    \end{figure*}
\subsection{Emission line fitting of ERIS datacube}
\label{sec:fitting_ERIS}

Figure~\ref{fig:int_spectrum} shows an integrated ERIS spectrum, extracted from an aperture of radius $0.5''$ centred on the position of the continuum peak, covering the $1.075-1.310$~\micron\ rest-frame spectral range. 
The spectrum beyond 1.31~\micron\ is not presented due to significant atmospheric absorption 
in this wavelength range.
We identify the strongest emission lines:
\HeI$\lambda1.083$\micron, Pa$\gamma$ $\lambda1.094$\micron, \PII $\lambda1.147$\micron, \PII$\lambda1.189$\micron, \FeII$\lambda1.257$\micron, and  Pa$\beta$ $\lambda1.282$\micron.
 The blue spectral windows in Fig.~\ref{fig:int_spectrum} indicate spectral regions of poor atmospheric transmission. This strongly affects the \Pag\ line and partially the \Pab\ line.


We fit emission lines in the ERIS datacube to derive the properties of the ionised gas in each spaxel.
In order to increase the signal-to-noise ratio (S/N), for each spaxel we create a mean spectrum by calculating the average of the spectrum from the selected spaxel and the eight adjacent spaxels (i.e. a $3\times3$ box). 
Then, we fit and subtract the continuum emission in each spaxel.
We perform a linear fit to the continuum across the entire wavelength range, excluding the spectral regions within 0.008~\micron\ from the position of the emission lines and regions of poor atmospheric transmission ($1.090-1.141$~\micron, $>1.29$~\micron\ rest-frame). 
The absence of stellar absorption lines in the ERIS spectrum renders the fitting of stellar continuum templates unnecessary. 

After subtracting the continuum, we model all identified lines simultaneously, using a combination of three Gaussian components. We constrain the kinematic parameters (velocity centroid and velocity dispersion) of each Gaussian component to be the same in all lines. We chose to use three components as it is the minimum number of components necessary to provide a reasonably good fit.
In particular, from the inspection of the line profile of the strongest lines (\HeI, \FeII, \Pab) across the field of view, we note the presence of a prominent blueshifted component, in addition to the narrow systemic component (see  spectrum in Fig.~\ref{fig:int_spectrum}). To model the line profiles, we also add a second broad component, preferentially redshifted or located at the same velocity as the narrow component, depending on the position in field of view. We stress that this component is not related to the BLR, as it is detected also in the forbidden \FeII\ transition. 
We checked that two Gaussian components are not enough to reproduce the asymmetric line shape, leaving significant structured residuals, and that using four Gaussian components does not produce a significant improvement in the fitting residuals, but introduce additional degeneracies. 

We allow the parameters to vary within predefined ranges, as summarized in Table~\ref{tab:fit_param}. The ranges were selected based on a visual inspection of the emission line profiles.
To retrieve the intrinsic line width, during the fit we convolve the Gaussian profiles with the instrument resolution ($\sim 25$~\kms, R$\sim 5000$).
We fit the emission lines using the Python non-linear least-squares minimisation routine ‘scipy.optimize.curve$\_$fit’.
The initial values of the parameters are specified in Table~\ref{tab:fit_param}. 
An example of the fit is shown in Fig.~\ref{fig:int_spectrum} for the total spectrum.
In the analysis, we mainly use parameters derived from the total line profile (e.g. percentile velocities) which are not dependant on the exact separation of the line profile in individual components. Thus, the final results are not strongly influenced by the initial values and ranges of the parameters. 

We produce maps of the percentile velocities $v10$, $v50$, and $v90$ - representing the velocities below which 10\%, 50\% and 90\% of the total line flux is located - calculated in each spaxel from the total best-fit profile. 
Before doing this, we excluded individual components with low S/N (peak S/N $<1.2$), to prevent them from biasing the velocity measurements.
Velocity maps are computed with respect to the systemic redshift $z= 0.01677\pm0.00001$, measured from the centroid of the narrow component in the fit of the total spectrum shown in Fig.~\ref{fig:int_spectrum}.
The maps of line fluxes and kinematics are presented in Sec.~\ref{sec:emission_line_maps}.

\subsection{Emission line fitting of X-shooter spectrum}
\label{sec:xshooter_fit}

In this section we describe the analysis of the X-shooter spectrum.
We are interested in fitting the main optical emission lines, 
in particular to derive the electron density from the \SII$\lambda6716/\lambda6731$\AA\ doublet ratio and the optical dust attenuation $A_V$ from the Balmer decrement, \Ha/\Hb.

We use a similar approach to the one applied for the ERIS spectra and described in the previous section (Sec.~\ref{sec:fitting_ERIS}). We consider the wavelength range $4800- 7000$~\AA. 
Specifically, we fit \Hb, \OIII$\lambda \lambda$4959,5007, \HeI$\lambda$5877, \OI$\lambda$6300,  \Ha, \NII$\lambda \lambda$6548,6583,  \HeI$\lambda$6678, and \SII$\lambda \lambda$6716,6731.
We fix the ratio of the \NII\ and \OIII\ doublet lines to the theoretical value of 2.99 \citep{Osterbrock2006}.
We let the \SII\ doublet line ratio free  to vary in the range $0.438 < $ \SII$\lambda6716$/\SII$\lambda6731 < 1.448$, which correspond to the theoretical limits for densities in the range $1-10^5$~cm$^{-3}$ \citep{Sanders2016a}.
Before fitting the emission lines, we model and subtract the continuum by interpolating the flux across the entire wavelength range, excluding the spectral regions within $\pm50$~\AA\ from the position of the emission lines. We use three kinematic components to model the emission lines, as  done for the ERIS spectra: a narrow component (n) and two broad components (b1 and b2).
We assume that all emission lines share the same kinematic parameters (velocity shift and velocity dispersion) for each corresponding Gaussian component.  
We use the Markov Chain Monte Carlo (MCMC) fitting code \emcee\ \citep{Foreman-Mackey2013} to sample the posterior distribution and derive the uncertainties of the parameters.
We use 60 walkers and 2000 iterations, which are enough to achieve convergence \citep[effective sample size > 10,][]{Gelman2004}. We consider the 50th percentile of the posterior distribution as the best-fit parameter, and we estimate the $1\sigma$ uncertainties as the difference between the 50th percentile and the 16th and 84th percentiles.
In Fig.~\ref{fig:spectra_opt} we show the best-fit model of the strongest emission lines 
(\Ha, \Hb, and \OIII, \NII, and \SII\ doublets). 

 We note that \MCG\ has previously been classified as a Seyfert~1.8 \citep{Oh2022}, 
indicating the presence of a BLR component in  \Ha\ and a weak BLR component in \Hb. 
However, our modelling does not require a BLR component, as the broad components that fit the \OIII\ line profile are sufficient to also reproduce the \Ha\ and \Hb\ profiles.
\citet{Oh2022} analysed the same X-shooter spectrum as in this work, but they model \Ha\ and \Hb\ with a narrow  and a BLR component (but no outflow), even though they also identified a prominent broad blueshifted component in \OIII. 
In contrast, we model each line with three Gaussians  (narrow+b1+b2). We also tested the inclusion of a BLR component (with FWHM $> 1000$~\kms), in addition to the other three Gaussian components, in our fit of \Ha\ and \Hb, but its best-fit amplitudes were negligible and its inclusion does not statistically improve the fit according to the Bayesian information criterion \citep[BIC,][]{Schwarz1978}. 

One of the goals of the fit is to measure the electron density from the ratio of the \SII\ lines. Figure~\ref{fig:SII_ratio} shows a zoom-in on the \SII\ doublet.
The two broad components of the \SII\ lines (b1 and b2) are blended, resulting in a strong degeneracy between them.
 The presence of the weak \HeI$\lambda6678$ line introduces additional uncertainty in the fit. To mitigate this, we compute the electron density for the `total broad component' by summing the two broad components. The posterior distribution of the \SII\ line ratio for the combined broad components (b1+b2) is presented in Fig.~\ref{fig:SII_ratio}.
We convert the line ratios to electron densities using the prescriptions from \citet{Sanders2016a}.
For the narrow component, we estimate an electron density $n_e= 1100^{+200}_{-160}$~cm$^{-3}$, while for the combined broad components, we find $n_e= 950^{+450}_{-250}$~cm$^{-3}$. These two measurements are consistent within the uncertainties. The electron density derived from the total line profile is $1000^{+200}_{-150}$~cm$^{-3}$. 

We derive the dust attenuation $A_V$ from the ratio of the \Ha\ and \Hb\ emission lines. Assuming Case B recombination, a temperature $T = 10^4$~K, 
and the \citet{Cardelli1989} extinction law, we derive $A_V(n)=1.78_{-0.25}^{+0.28}$ for the narrow component, $A_V(b1)=0.28_{-0.28}^{+0.52}$ for the broad blueshifted component and $A_V(b2)=0.40_{-0.30}^{+0.32}$ for the second broad component. For the sum of the two broad components, we obtain $A_V(b1+b2)=0.35_{-0.35}^{+0.68}$. This suggests a lower dust attenuation in the outflow relative to the narrow, galaxy disk component, consistent with previous observations of local Seyfert galaxies \citep[e.g.][]{Mingozzi2019}.
The properties derived from the fit of the X-shooter spectrum are summarized in Tab.~\ref{tab:optical_prop}.

\section{Results}
\label{sec:results}

\subsection{Comparison of peaks positions in the different bands}
The middle panel of Fig.~\ref{fig:nuclei_position} shows the \Pab\ best-fit integrated flux map from ERIS, overplotted with the positions of the nuclei detected at different wavelengths: the three \OIII\ peaks detected by HST 
and the two X-ray positions that we derived from the  \FeKa\ line map. 
We also overlay the contours of the VLA 33 GHz radio data, that will be discussed in Sec.~\ref{sec:radio}.

The \Pab\ flux follows in broad terms the \OIII\ morphology (see Fig.~2 in \citealt{TrindadeFalcao2024}).
In the north, there is a \Pab\ peak coincident with the \OIII\ peak, the X-ray and the VLA positions. This was identified as the `north nucleus' by \citet{TrindadeFalcao2024}.
A second peak of \Pab\ flux is located close to the middle \OIII\ peak and the continuum position. 
In the south, the \Pab\ emission extends toward the location of the south \OIII\ peak. 
The south X-ray peak is in agreement with the position of the south \OIII\ peak, suggesting the second AGN could be located there. The central \OIII\ peak is close to the location of one of the radio spots.  We will discuss further the location of the nuclei and the evidence for a dual AGN  in Sec.~\ref{sec:discussion}.

\subsection{Emission line maps and kinematics}
\label{sec:emission_line_maps}

\begin{figure*}[!t]
\centering 
\includegraphics[width=0.9\textwidth, trim=0mm 3mm 0mm 0mm, clip]{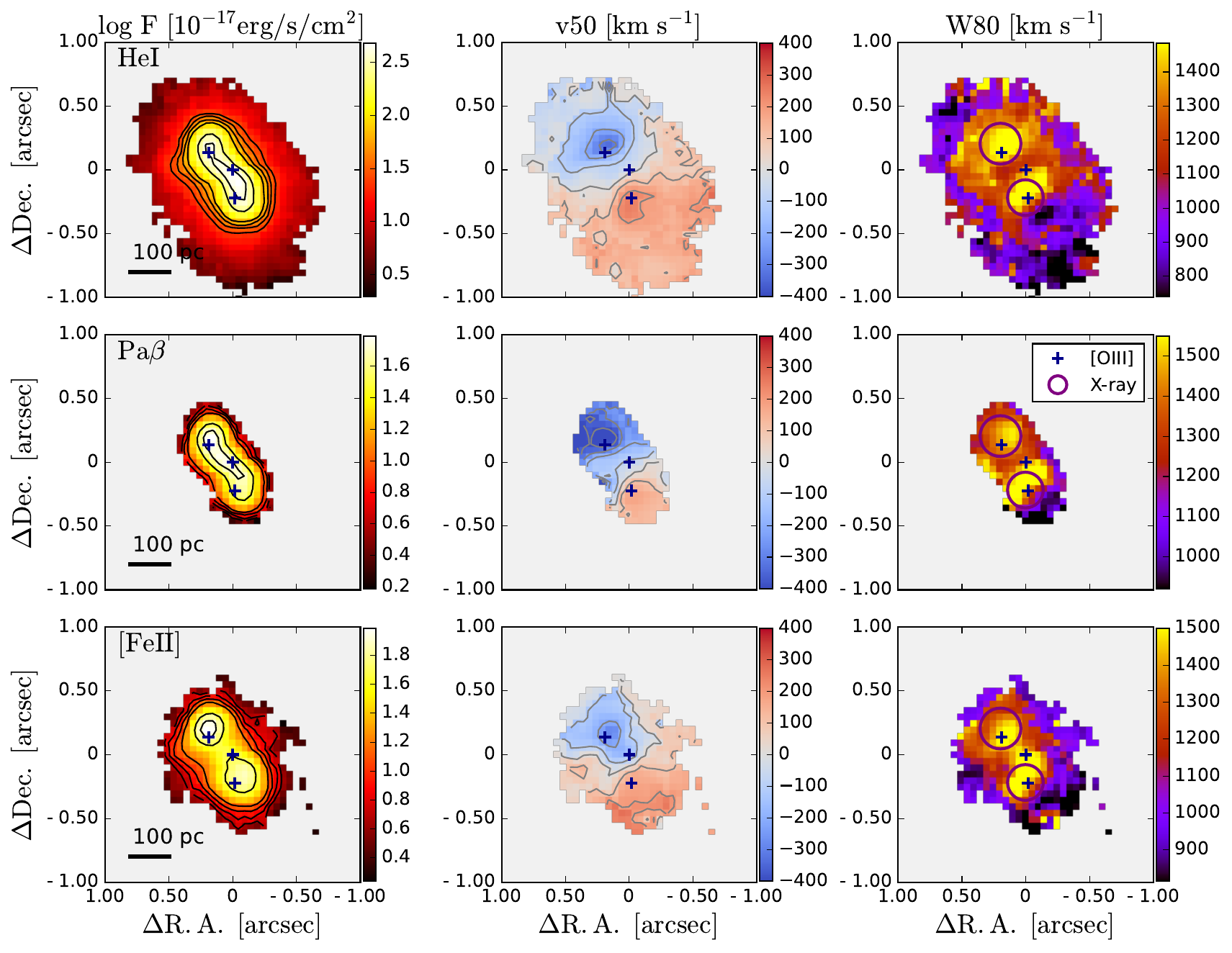}
\caption{Map of the total flux, average velocity ($v50$) and line width $W80$ (width
encompassing 80\% of the flux) derived from total line profile fitted with three Gaussian components. \textit{From top to bottom}: \HeI$\lambda1.083$~\micron,  \Pab,  \FeII$\lambda1.257$~\micron.
We show only spaxels with peak S/N in the total profile above 3.
The blue $`+$' symbols show the positions of the \OIII\ peaks; the purple circles show the position of the X-ray peaks. 
Velocity maps are shown with respect to the systemic redshift  $z= 0.01677$.
Contours in the first column show arbitrary flux levels, in the second column velocity contours are in intervals of 100~\kms. }
\label{fig:flux_maps}
\end{figure*}

\begin{figure*}[!h]
\centering 
\includegraphics[width=0.99\textwidth]
{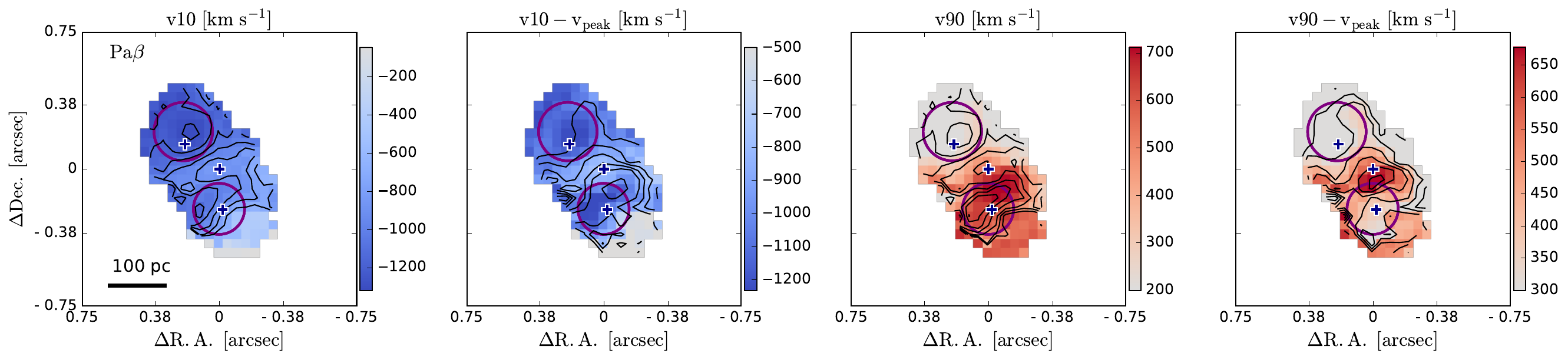}
\caption{Velocity maps of the \Pab\ emission line. From left to right: map of the percentiles velocities $v10$, the difference between $v10$ and the velocity at the peak of the line profile ($v_{peak}$), $v90$, and the difference $v90-v_{peak}$.
We show only spaxels with peak S/N in the total profile above 3.
The blue $`+$' symbols show the positions of the \OIII\ peaks; the purple circles show the position of the X-ray peaks.
Velocity maps are shown with respect to $z= 0.01677$.}
\label{fig:kinematic_maps}
\end{figure*}

In Fig.~\ref{fig:flux_maps} we show the maps of the flux and kinematics of the brightest emission lines detected in the ERIS data: \HeI, \Pab, and \FeII. The three columns show the total flux, median velocity ($v50$) and line width $W80=v90-v10$ (i.e. the width encompassing 80\% of the line flux) obtained from the fit with three components. 
The maps obtained from the individual Gaussian components are shown in Appendix~\ref{sec:appendix_kin_maps}.

The morphologies of the \Pab\ and \HeI\ flux maps are similar and follow the \OIII\ flux distribution:  the near-IR lines show an `S'-shape extending from the north to the south \OIII\ peaks \citep[see][]{TrindadeFalcao2024}. 
 The \FeII\ flux instead shows two peaks: the first peak is located $\sim0.1$\arcsec\ to the north of the northern \OIII\ peak; the second peak is located $\sim0.1$\arcsec\ to the north-west of the southern \OIII\ peak.

The $v50$ velocity maps of the three lines (middle column in Fig.~\ref{fig:flux_maps}) show an overall gradient increasing from the north-east to the south-west, spanning a range from approximately $-400$~\kms\ to $200$~\kms. The most blueshifted velocities are found in the position of the north \OIII\ peak,  while the most redshifted velocities are located at the position of the south \OIII\ peak. 
The velocity pattern of the ionised gas is consistent with the inner part of a disturbed disk, with the kinematic major axis approximately aligned with the morphological major-axis (see Fig.~\ref{fig:nuclei_position}). Alternatively, the velocity pattern could be explained by two circumnuclear components around the two nuclei, in the process of merging. 
Disturbed ionised gas kinematics are commonly observed in the central regions of infrared-luminous galaxy mergers, consistent with the presence of a perturbed disk on large scale (> 100 pc) or of circumnuclear components \citep[e.g.,][]{Perna2022, Ulivi2025, Ceci2025, Carlsen2025arXiv}. 

The third column shows the maps of the $W80$ line-width for the three emission lines. 
All three maps show two regions with high $W80$ (>1400~\kms). The region in the north is co-spatial with the north \OIII\ and X-ray peaks. 
The one in the south is roughly coincident with the position of the south \OIII\ and X-ray peaks. 
These two regions may indicate the presence of high-velocity outflows. We will discuss them in more detail in Sec.~\ref{sec:outflows}.

Figure~\ref{fig:kinematic_maps} shows the maps of the percentile velocities $v10$ and $v90$ of the \Pab\ line. The $v10$ map shows two peaks, which correspond roughly to the location of the two X-ray peaks. 
The $v90$ map also shows a peak  at the position of the north X-ray centroid with values of $\sim200$~\kms, and another enhancement close to the south X-ray and \OIII\ centroids, reaching $\sim640$~\kms. 
These maps support the interpretation of the $W80$ enhancements being related to two outflows originating from the X-ray positions.
In Fig.~\ref{fig:kinematic_maps} we also show maps of the difference between $v10$ and the velocity at the peak of the total line profile \citep[$v_{\rm peak}$, e.g.][]{Cresci2015}. In this way, we remove the effect of possible rotation, assuming that $v_{\rm peak}$ is tracing the systemic gas. 
This map shows velocities of around $-1200$~\kms\ close to the positions of the two X-ray peaks. For the redshifted part, we produce a map of $v90-v_{\rm peak}$. In this map, we see a peak of $\sim700$~\kms\ close to the south X-ray peak, possibly indicating the red part of the outflow.

\subsection{Outflow properties}
\label{sec:outflows}

\begin{table*}
\centering
\caption{Outflow properties derived from the \Pab\ line.}
\setlength{\tabcolsep}{3pt}
\begin{tabular}{lccc|ccccc|cccc|ccc}
\hline
Region & $\log L_{bol}$ & \multicolumn{2}{c|}{$R_{out}$} & $v10$ or $v90$ & method &  $M_{out}$ & \Mrate  & $\log \dot{E}_{out}$   & $v_{max}$  & $v02$ \\ %
        & [erg s$^{-1}$] & [\arcsec] & [pc] &  [\kms]   & & [$\times 10^4$ \Msun] & [\Msun\ yr$^{-1}$] & [erg s$^{-1}$] & [\kms] & [\kms]& \\
(1) & (2)  & (3) & (4) &(5) & (6) & (7) & (8) & (9) & (10)& (11)\\ 
  \hline \hline

north & $44.2$ & 0.2 & $67$ &$-1210$  & $v<-300$ & $35\pm8$ & $20\pm5$ &  $42.9$  &  $-1780\pm30$ & $-1700$\\
 & & & &  & blue comp.  & $38\pm9$ & $21\pm6$ & $43.0$ & \\
 \cline{5-9}  
  &  & &  & 218 & $v>300$ & $5\pm1$ & $0.5\pm0.1$ &  $39.9$  &   & \\
 & & & &  & red comp.  & $4\pm2$ & $0.5\pm0.2$ & $39.8$ & \\
 \hline
south &$44.0$ & 0.2 & $67$ & $-1140$ & $v<-300$  &  $36\pm9$ & $20\pm5$ &  $42.9$ & $-1610\pm50$ & $-1680$\\
& & & & & blue comp.  &  $32\pm8$ & $17\pm5$ &  $42.9$ &   \\
 \cline{5-9}  
& & & & 270 & $v>300$  &  $6\pm2$ & $0.8\pm0.2$ &  $40.8$ &   \\
& & & & & red comp.  &  $14\pm4$ & $1.7\pm0.6$ &  $40.6$ &   \\
\hline
total & $44.4$ &0.5 & 167 & $-1100$ & $v<-300$   &  $84\pm20$ & $17\pm4$  & 42.8 & $ -1730\pm40$ & $-1690$\\ 
& & & & & blue comp. &$74\pm19$ & $15\pm4$  & 42.8 \\
 \cline{5-9}  
& & &  & $420$ & $v>300$   &  $29\pm7$ & $2.2\pm0.5$  & 41.1 &  & \\ 
& & & & & red comp. &  $22\pm9$ & $1.7\pm0.7$  & 41.0 & &\\
\hline

\end{tabular} 
\label{tab:outflow_prop}
\tablefoot{
(1) Regions from which the spectra have been extracted. The `total' spectrum is from a region of 0.5\arcsec\ radius centred on the continuum peak, that encompasses the two nuclei.
(2) Bolometric luminosity estimated from the \OIII\ luminosity, taken from \citet{TrindadeFalcao2024}.
(3) and (4) Radius of the outflow, equivalent to radius of the aperture used to extract the spectrum,
in arcsec and pc, respectively.
(5) 10th (if negative) or 90th (if positive) percentile velocity.
(6) Method used to calculate the outflow mass (see Sec.~\ref{sec:outflows}).
(7) Outflow gas mass estimated from the \Pab\ line using the flux of the broad blueshifted component and equation~\ref{eq:Mout}, assuming an electron density $n_e=1100\pm250$~cm$^{-3}$.
(8) Mass outflow rate, estimated as \Mrate$= 3\times M_{out}\times v10/R_{out}$.
(9) Outflow kinetic power, defined as $\dot{E}_{out}=\frac{1}{2}\times\dot{M}_{out}\times v_{out}^2$.
(10) Maximum outflow velocity, derived from a fit with two Gaussian components and defined as $v_{max}= (v_{broad}-v_{narrow})-2\times \sigma_{broad}$,  following \citet{Fiore2017}.
(11) 2nd percentile velocity.
}
\end{table*}

\begin{figure}[!t]
\centering 
\includegraphics[width=0.4\textwidth]{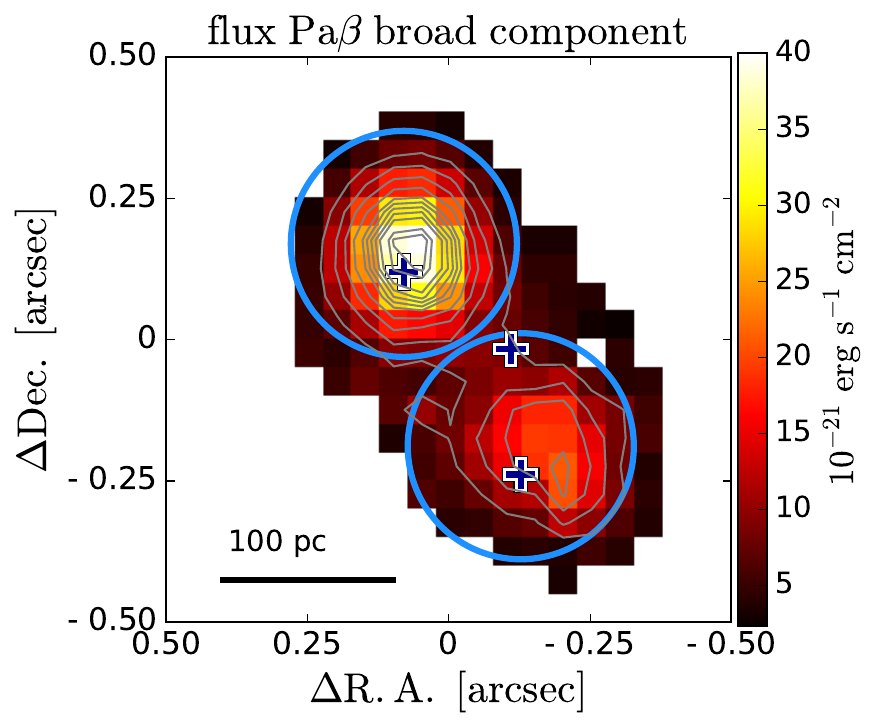}
\includegraphics[width=0.44\textwidth]{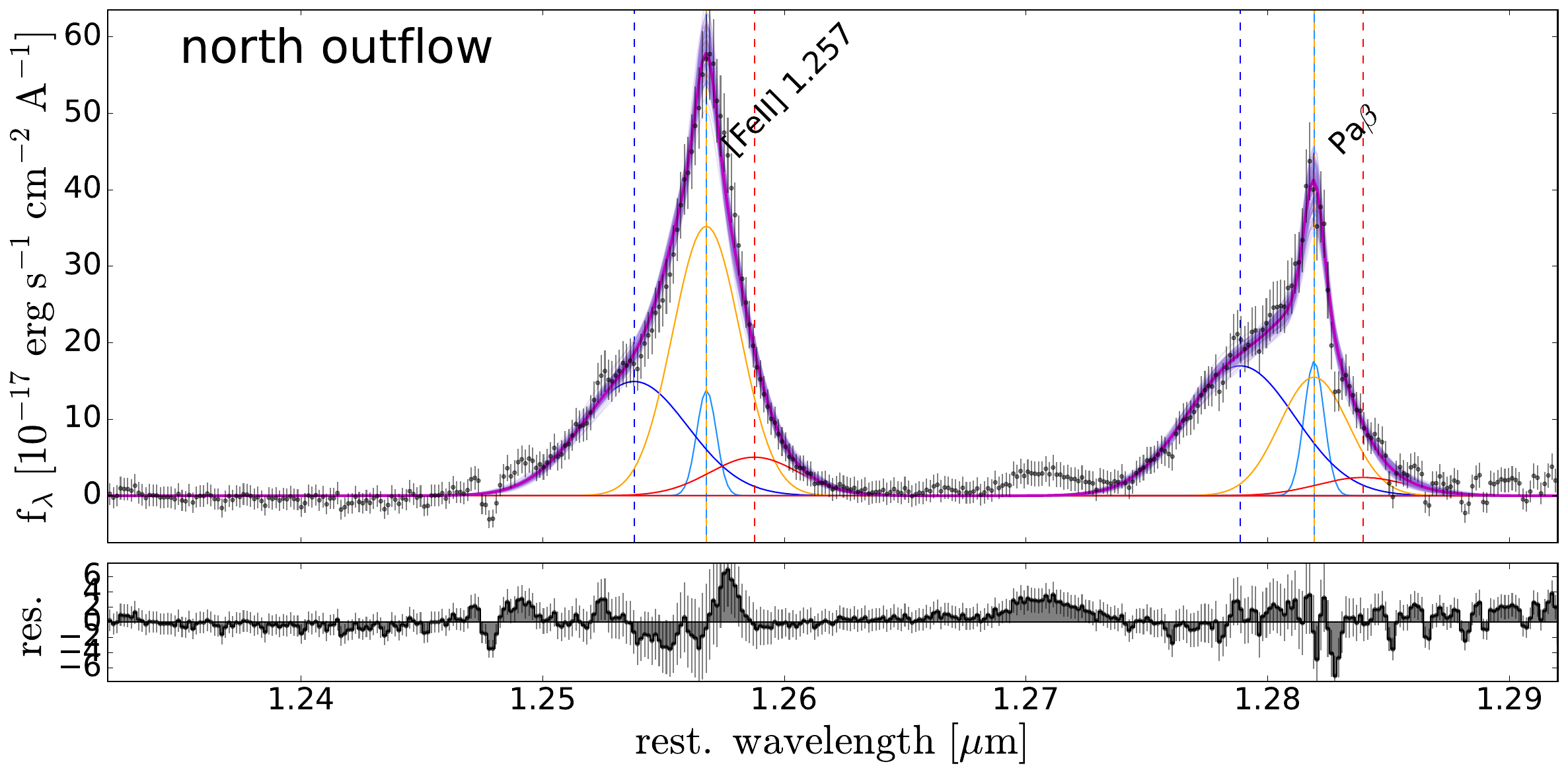}
\includegraphics[width=0.44\textwidth]{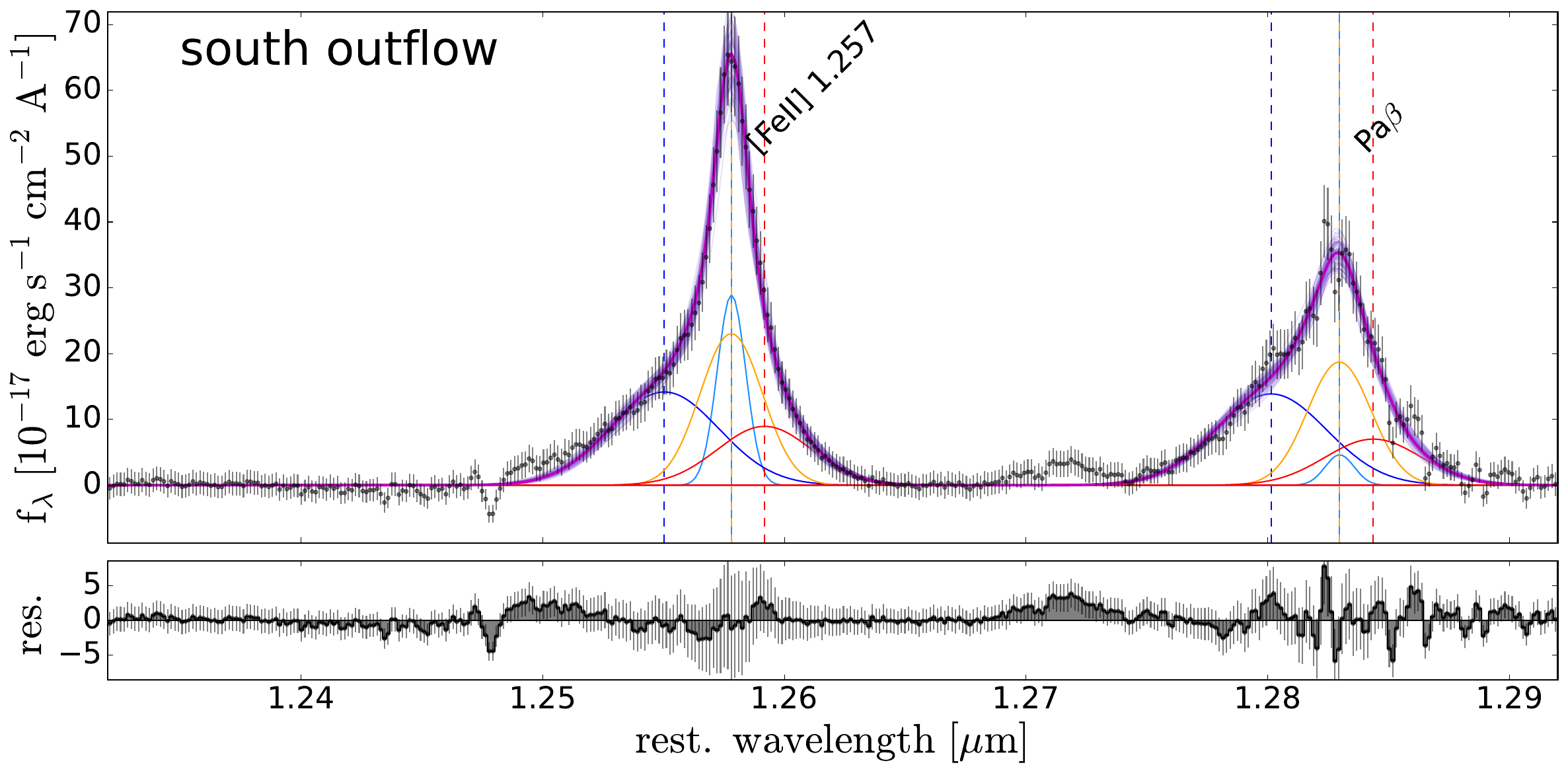}
\caption{\textit{Top panel:} Map of the broad blueshifted (b1) component of \Pab. Only spaxels with S/N>2 are shown. Grey contours show arbitrary flux levels. The two lightblue circles show the regions used to extract the outflow spectra, with radii of 0.20\arcsec. Blue crosses show the \OIII\ peaks. 
 \textit{Middle and bottom panels:} Spectra extracted from the north and south outflow regions. 
Data are shown as black dots with 1$\sigma$ uncertainties in grey. The emission lines are fitted with four Gaussian components: narrow (lightblue), broad systemic (orange), broad blueshifted (blue) and  broad redshifted (red). The dashed vertical lines show the centroid of the individual components. The total best-fit model is shown in magenta, the uncertainties on the model are illustrated by multiple MCMC realizations shown in violet.} 
\label{fig:spectra_outflow}
\end{figure}

As discussed in the previous section, we identify two regions with enhanced line width $W80$, which correspond approximately to the positions of the north and south \OIII\ peaks and the two X-ray peaks. 
In Sec.~\ref{sec:discussion} we will discuss the possible origin of these two outflow components, in particular whether they originate from two AGN or from a single source. In this section, we characterize the properties of the two outflows, assuming they originate from two nuclei. If they originate from a single AGN, the total mass of the outflow will be the sum of the north and south components. 
The top panel of Fig.~\ref{fig:spectra_outflow} shows the flux distribution of the broad blueshifted component of \Pab, which exhibits two clear peaks. 
The outflows are not spatially resolved, thus preventing a proper geometry and orientation determination. 
 Therefore, in this section we report only the integrated properties of the two outflows. 
 To derive the outflows properties, we extracted  spectra from two apertures of 0.2\arcsec\ radius, centred on the two outflow regions, defined from the map of the broad blueshifted \Pab\ component (see circles in Fig.~\ref{fig:spectra_outflow}). These apertures include the totality of the emission of the two outflow regions.
We will refer to these two spectra, shown in Fig.~\ref{fig:spectra_outflow}, as the `outflow spectra' hereafter. 

 We calculated the outflow properties (mass, velocity and radius) for each outflow, independently.
 We choose to focus on the \Pab\ line as converting the line luminosity of an hydrogen line to a gas mass does not require any assumption on the elemental abundances and hardness of the radiation field \citep[e.g.,][]{Carniani2015, Baron2019, Venturi2023}. 
 Both outflow spectra show a prominent blueshifted wing. 
We fit the two spectra using four Gaussian components, with a similar method as the one used for the spaxel-by-spaxel fitting (see Sec.~\ref{sec:fitting_ERIS}).
We use two component to trace the systemic gas (one narrow and a broader component), and two components to model the blue and redshifted part of outflows. This approach could not be used for the spaxel-by-spaxel fit due to the lower S/N. We note that the derived outflows parameters (outflow velocity and mass, as explained below) do not strongly depend on the separation of the line profile in different components, but just need an overall good modelling of the total line profile.

We use the MCMC code \emcee\ to fit the lines and estimate the uncertainties. The details of the MCMC fit are explained in Sec.~\ref{sec:xshooter_fit}.
The best-fit models are shown in Fig.~\ref{fig:spectra_outflow}.
In the following, we describe how we derive the different outflow properties, which are summarized in Table~\ref{tab:outflow_prop}.
\vspace{-1\baselineskip}
\paragraph{\textit{Radius:}}
We adopt the aperture radii (0.2\arcsec) as the projected radii of the outflows, corresponding to $\sim67$~pc.  
This value approximately matches the spatial extent of the broad blueshifted component (see Fig.~\ref{fig:spectra_outflow}).
We note that we are only considering the projected radii, as  the inclination of the outflows in unknown. Thus, the reported radii should be considered as lower limits of the true outflow sizes. 

\vspace{-1\baselineskip}
\paragraph{\textit{Velocity:}}
We consider as the outflow velocities the $v10$ percentile velocities estimated from the best-fit model of the \Pab\ line \citep[following][]{Cresci2015, Harrison2016, Tozzi2024, Bertola2025}. The assumption behind this approach is that the outflow broadens the line profile because it has multiple components in multiple directions, and the maximum velocity in the blushifted wings is produced by the components closer to the light of sight, which are less affected by projection effects. This scenario assumes that the outflow velocity is purely radial and constant. 
For each outflow, we measure the $v10$ velocity relative to the `systemic' velocity, defined for each spectrum as the centroid of its narrow Gaussian component. 
For the north and south outflows, we find $v10=-1210$~\kms\ and $v10=-1140$~\kms, respectively.
We also estimate the outflow velocity using the definition $v_{\rm max} = |v_{\rm broad}-v_{\rm narrow}|+2\cdot\sigma_{broad}$, where $v_{\rm narrow}$ and $v_{\rm broad}$ are the velocity shifts of the narrow and broad components of a two-component fit, and  $\sigma_{\rm broad}$ is the velocity dispersion of the broad  component \citep{Rupke2013}. To estimate these parameters, we performed a new fit using the same method as explained before, but with two Gaussian components (narrow + broad) instead of four. We obtained $v_{\rm max}=-1780$~\kms\ and $v_{\rm max}=-1610$~\kms, for the north and south outflows, respectively.
We prefer to use the $v10$ parameter, as it is less dependent on the model assumed for the fitting, but we also consider this definition as it is often used in the literature \citep[e.g.][]{Fiore2017,Rojas2020, Musiimenta2023}.
For the redshifted part of the outflow, we consider the $v90$ velocity.

\vspace{-1\baselineskip}
\paragraph{\textit{Mass:}}
We estimate the outflow mass from the flux of the \Pab\ line using two different methods. We consider as outflow flux: 1) all the flux above a given velocity threshold ($|v|>300$~\kms), and 2) the flux belonging to the broad blueshifted (or redshifted) Gaussian components. We select the threshold of $|v|>300$~\kms\ as rotating gas typically does not reach velocities above this threshold, and this is the typical velocity at which the broad blueshifted component starts to dominate over the narrow component in our outflow spectra. 
We prefer the first  method, as it is 
less model dependent \citep[e.g., ][]{Harrison2014, Kakkad2020}. However, we report the results of both methods in Table~\ref{tab:outflow_prop}.
We note that the red part of the outflow is fainter probably due to the obscuration by the host galaxy, resulting into smaller outflow masses.

We corrected the \Pab\ luminosity for attenuation using the value $A_V(b)=0.28$, derived from the ratio of the broad component fluxes of \Ha\ and \Hb\ assuming the \citet{Cardelli1989} extinction law (see Sec.~\ref{sec:xshooter_fit}). We prefer to use \Ha/\Hb\ over \Ha/\Pab\ to avoid possible flux calibration issues and problems due to the difference in the slit sizes between the X-shooter and the ERIS data. Given the low value of $A_V$, the correction for the \Pab\ fluxes is only a factor of 1.07.
We converted the extinction-corrected \Pab\ luminosity to \Ha\ luminosity, assuming the theoretical line ratio \Ha/\Pab =17.57 \cite[Case B recombination, $T = 10^4$~K,][]{Draine2011_book}. Then, we computed the outflow mass following \citet{Cresci2017}:
\begin{equation}
\label{eq:Mout}
    M_{\rm out} = 3.2 \times  10^5 \times \frac{L_{\rm H\alpha}}{10^{40}~\text{erg/s}} \times  \frac{100~\text{cm}^{-3}}{n_e}  M_\odot,
\end{equation}

\noindent
where $n_e$ is the electron density and $L_{\rm H\alpha}$ is the \Ha\ line luminosity. We estimated the electron density from the \SII $\lambda \lambda$6716,31 line ratio, using the X-shooter spectrum (see Sec.~\ref{sec:xshooter_fit}). The mean electron densities derived for the narrow and broad components are in the range $950-1100$~cm$^{-3}$, while the total profile gives $1000\pm200$~cm$^{-3}$. Since the measurements in the different components are consistent within the uncertainties, we assume the electron density derived from the total profile for our calculation.
Considering our preferred method (flux at $|v|>300$~\kms), we derived outflow masses of  $(40\pm8)\times 10^4$~\Msun\ and $(42\pm8)\times 10^4$~\Msun\ for the north and south outflows, respectively.
The masses derived using the second method are consistent with these values within the uncertainties.

\vspace{-1\baselineskip}
\paragraph{\textit{Mass outflow rate:} }
Following \citet{Cresci2015} and \citet{Fiore2017}, we estimated mass outflow rates as
\Mrate$= 3\times M_{\rm out}\times v_{\rm out}/R_{\rm out}$.
 For our preferred method, we find 
 $20\pm5$~\Msunyr\ for the blueshifted part and 0.5-0.8~\Msunyr\ for the redshifted part of each outflow.

\subsection{Comparison of outflow properties with the literature}
\label{sec:outflow_literature}

\begin{figure}[!t]
\centering 
\includegraphics[width=0.45\textwidth]{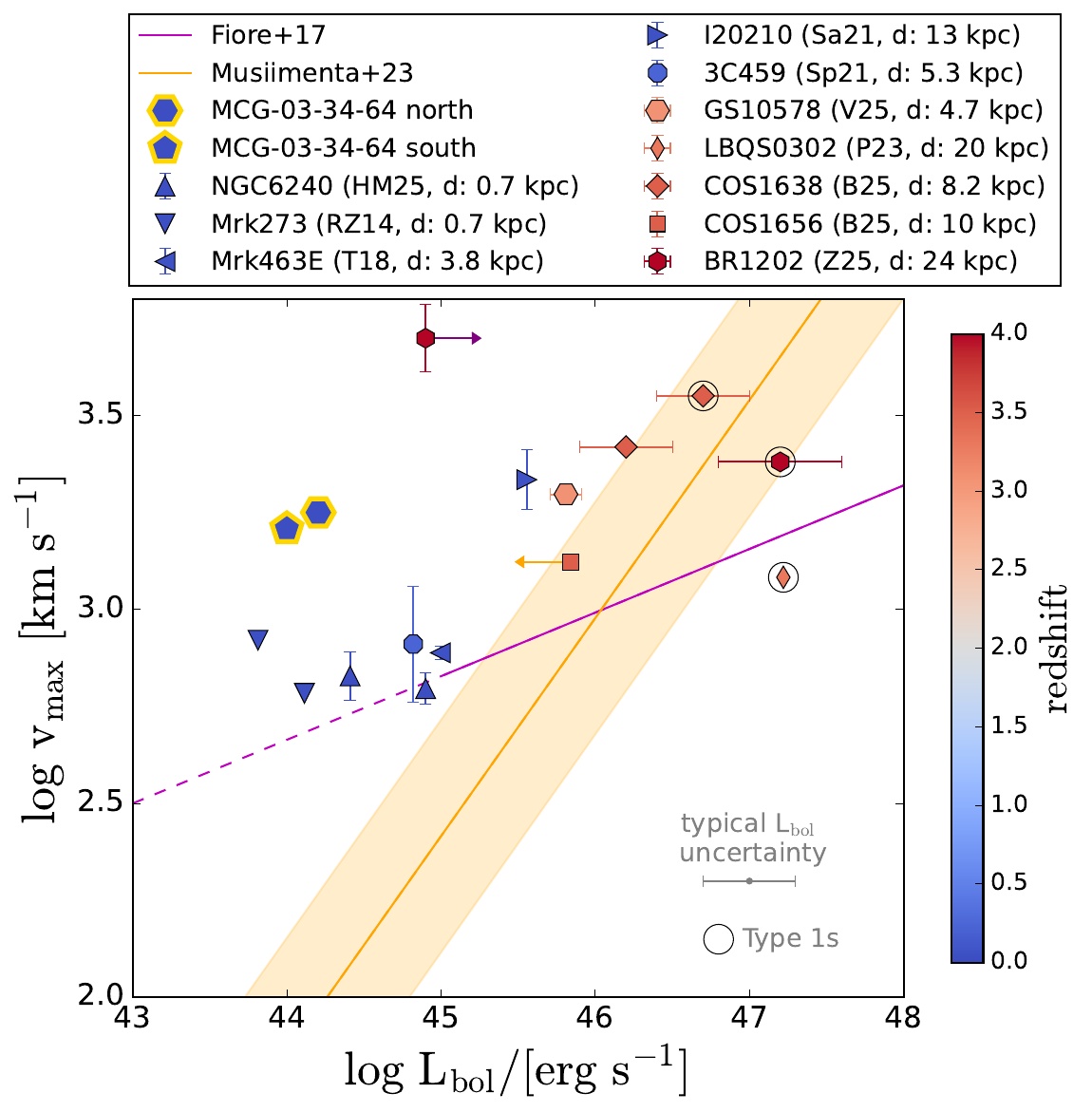
}

\caption{Bolometric luminosity versus outflow velocity for dual AGN (candidates) from the literature. The magenta line shows the \citet{Fiore2017} relation, the orange line and shaded area show the \citet{Musiimenta2023} relation and its 1~$\sigma$ scatter. 
Dual AGN are color-coded according to their redshift (HM25 \citealt{Hermosa-Munoz2025};
RZ14 \citealt{RodriguezZaurin2014};
T18 \citealt{Treister2018}; 
Sa21 \citealt{Saturni2021};
Sp21 \citealt{Speranza2021};
V25 \citealt{Venturi2026arXiv};
P23 \citealt{Perna2023}; B25 \citealt{Bertola2025};
Z25 \citealt{Zamora2025}).
The north and south outflows of \MCG\ are highlighted with a yellow border.
Type 1 AGN are indicated with grey circles. The typical uncertainties on $L_{\rm bol}$ of 0.3~dex is indicated in the bottom right.} 
\label{fig:Lbol_vs_vmax}
\end{figure}

\begin{figure*}[!h]
\centering 
\includegraphics[width=0.99\textwidth, trim=0mm 3mm 0mm 0mm, clip]
{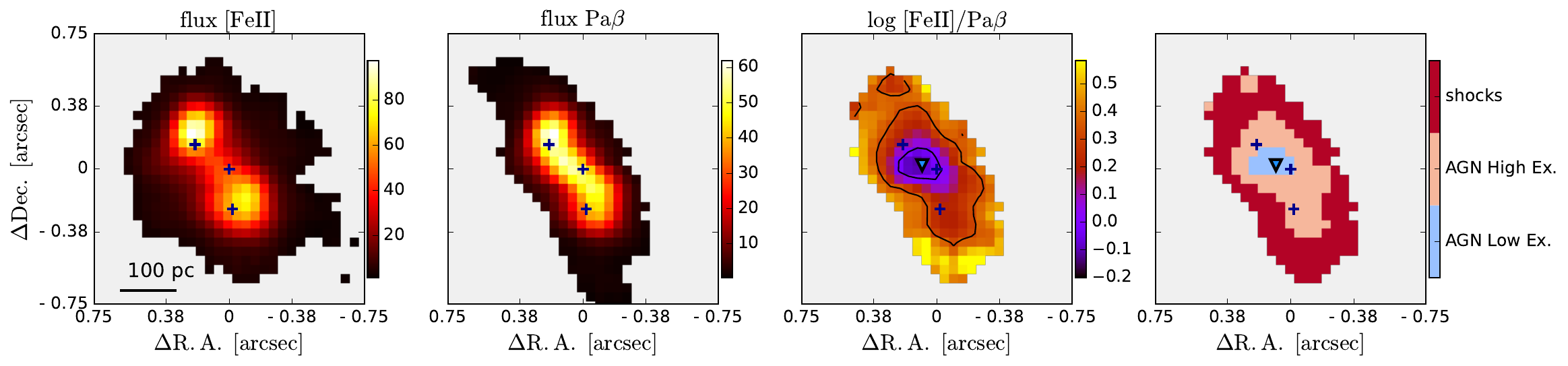}
\caption{Comparison of \FeII\ and \Pab\ total fluxes.   From left to right:  the \FeII\ total flux map, \Pab\ total flux, \FeII\ over \Pab\ flux ratio map, diagnostic map based on the \FeII/\Pab\ ratio. Only spaxels with S/N> 3 are shown. The right panel shows the spaxels dominated by  AGN low and high excitation and shocks,  according to the diagram from \citet{Bianchin2024} 
None of the spaxels are dominated by star formation. The separation are shown with black contours on the line ratio map.  Symbols as in Fig.~\ref{fig:nuclei_position}.}
\label{fig:map_ratio_FeII_Pab}
\end{figure*}

\begin{figure*}[!h]
\centering 
\includegraphics[width=0.7\textwidth, trim=0mm 3mm 0mm 0mm, clip]
{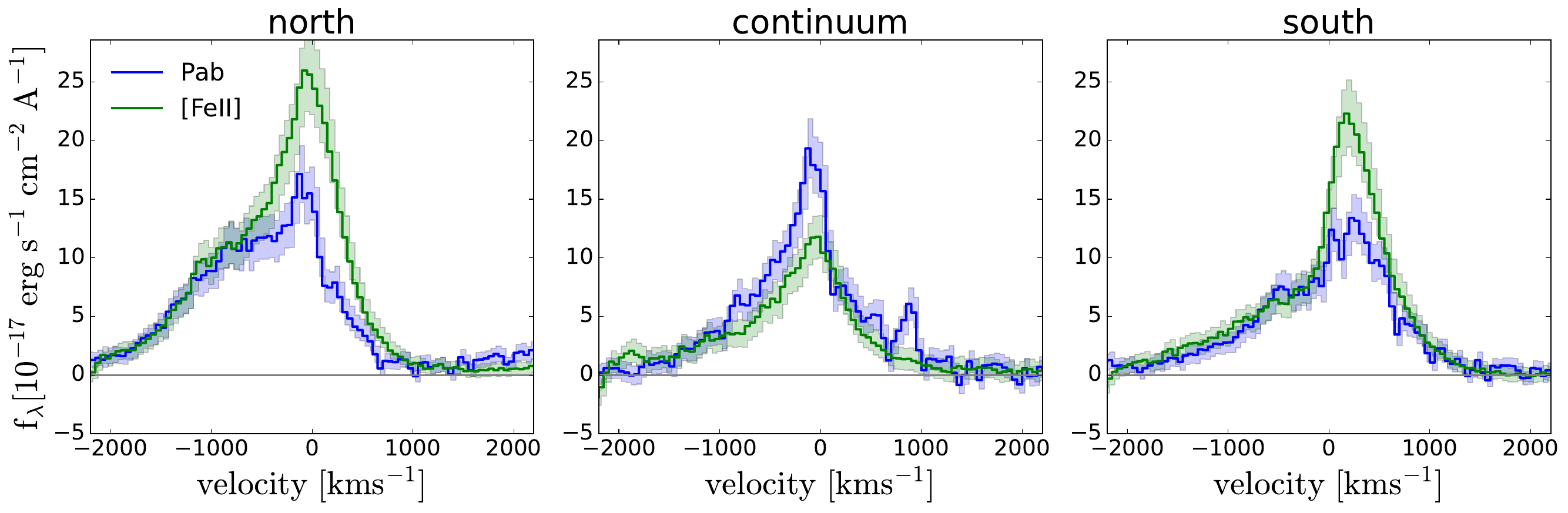}
\includegraphics[width=0.27\textwidth, trim=0mm 3mm 0mm 0mm, clip]
{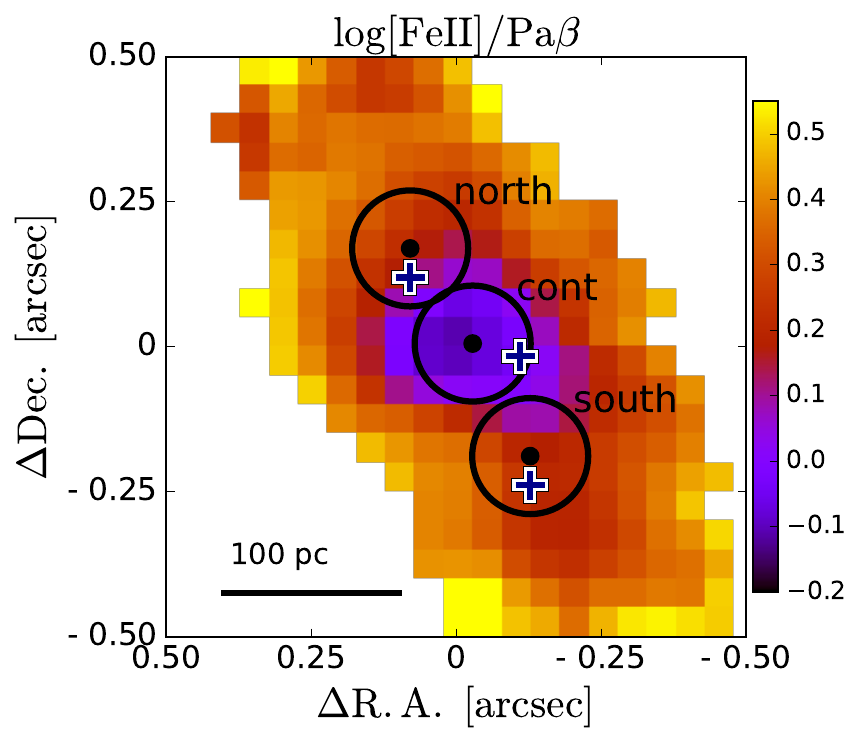}
\caption{Comparison of \FeII\ (green) and \Pab\ (blue) line profiles in the three regions shown on the right panel (extraction radius 0.1\arcsec): north \OIII\ peak, continuum peak, and south \OIII\ peak. In the spectrum extracted from the continuum peak, which is the region with lowest \FeII /\Pab, the \Pab\ profile shows an excess of narrow emission at velocities close to zero with respect to \FeII.}
\label{fig:comp_FeII_Pab_profiles}
\end{figure*}

In this section, we compare the outflow properties of \MCG\ with other samples of nearby AGN from the literature. 
We assume that the north and south outflows are associated with the candidate north and south nuclei. Given the similar bolometric luminosities and outflow properties of the two nuclei, the conclusions do not change significantly if we were to assume that the two outflows were the two side of a single outflow produced by a single AGN. We discuss these two scenarios in Sec.~\ref{sec:discussion}.

We compare the outflow velocities of the two outflows of \MCG\ with a sample of $\sim600$ nearby ($z<0.2$) X-ray selected AGN from the Sloan Digital Sky Survey (SDSS) presented in \citet{Perna2017a}. They study the \OIII$\lambda5007$ emission line and parametrise the outflow velocities in terms of $v02$, i.e. the 2nd percentile velocity. 
The two outflows of \MCG\ have $v02\sim-1700$~\kms\ (similar to the $v_{max}$ values, see Tab.~\ref{tab:outflow_prop}), which corresponds to the top $5\%$ of the distribution of velocities in the sample by \citet{Perna2017a}.
If we consider only AGN with $\log L_{\text{[OIII]}}/[\text{erg s}^{-1}]=41-42$ and $z<0.2$, to better match the properties of \MCG, our target is in the top $3\%$ of the distribution.
We also compare \MCG\ with the sample of 180 nearby ($z<0.7$, median $<z>=0.06$) AGN from the BASS survey, whose ionised outflow properties have been studied using the \OIII$\lambda5007$ line in  \citet{Rojas2020}.\footnote{\MCG\ is also in the sample by \citet{Rojas2020}. They find $v_{\rm max}=-1070^{+990}_{-80}$~\kms, lower, but still in agreement within the uncertainties with our values of $v_{\rm max}=[-1600,-1800]$~\kms\ for the two nuclei. From the fit of the \OIII\ line in the X-shooter spectrum, we find $v_{\rm max}=1550\pm5$~\kms. Our uncertainties are smaller because we use a higher quality spectrum. The spectrum used in \citet{Rojas2020} was taken with the SAAO~1.9m telescope, with a 2\arcsec\ slit.} 
The AGN in this sample have $\log L_{\text{[OIII]}}/[\text{erg s}^{-1}]=39.1-43.7$ and  $\log L_{\rm bol}/[\text{erg s}^{-1}]=42.7-47.5$.
Comparing the $v02$ velocities, \MCG\ is in the top $1\%$ of the distribution. 
So, overall it seems that the outflows in \MCG\ have higher velocities compared with local AGN at similar bolometric luminosity. This could be related to the fact that \MCG\ hosts a dual AGN. 

To test this, we investigate whether fast outflows are also common in other dual AGN from the literature.
We collected dual AGN  
with available measurements of ionised outflow velocity and bolometric luminosity for each nucleus. The outflow velocity is defined as $v_{\rm out}= \Delta v_{\rm broad}+2\sigma_{\rm broad}$ for all targets, to compare with the scaling relations from the literature \citep{Fiore2017, Musiimenta2023}.
 In Table~\ref{tab:literature_outflows} we report the list of targets, their outflow velocities, bolometric luminosities, and the corresponding references.
In Figure~\ref{fig:Lbol_vs_vmax} we show the outflow velocity 
versus bolometric luminosity for \MCG\ and the dual AGN  from the literature.
Dual AGN tend to be on or above the relations from \citet{Fiore2017} and \citet{Musiimenta2023}. Thus, they seem to have velocities higher than the average for a given AGN bolometric luminosity. We note that the sample of dual AGN is not complete, and it could suffer from selection effects. For example, low-velocity outflows are more difficult to identify. Moreover, both the \citet{Fiore2017} and \citet{Musiimenta2023} samples have poor coverage at low luminosities ($\log L_{\rm bol}/[\text{erg s}^{-1}]<45$), and in general the scatter in this relation is very large at low luminosities.
We also note that the literature dual AGN have very different redshift (from z=0 to z=4) and nuclear separations (from $\sim20$~kpc to sub-kpc scale), and we check that the offset from the scaling relation of \citet{Fiore2017} does not correlate with nuclear separation or redshift.
Thus, even though we find a hint that dual AGN may have higher outflow velocities, there are many caveats.
The high outflow velocities in \MCG\ could also be due to other factors. \MCG\ is a LIRG, thus it is expected to be gas-rich in the central region and to have fast outflows \citep[e.g.][]{Arribas2014, Fluetsch2021}. Moreover, mergers may help to fuel gas towards the central regions, feeding the AGN and producing faster outflows. 
 Additionally, \MCG\ has a radio jet in its centre, that may contribute to push the outflow \citep[e.g.][]{Cresci2015a, Finlez2018, May2020}. 

We compare the mass-outflow rate derived for \MCG\ with the relation between \Mrate\ and $L_{\rm bol}$ reported in the literature for ionised outflows by \citet{Fiore2017} and \citet{Musiimenta2023}.
We consider the mass outflow rates calculated using the $|v|>300$~\kms\ method, but we note that the two methods agree within 10\% (see Table~\ref{tab:outflow_prop}). 
The bolometric luminosity estimated for the two nuclei from the \OIII\ luminosity are $ L_{\rm bol}=10^{44.2}$~erg~s$^{-1}$ (north) and $ L_{\rm bol}=10^{44}$~erg~s$^{-1}$  (south) \citep{TrindadeFalcao2024}.
Our mass-outflow rates are above the extrapolation to low $L_{\rm bol}$ of the \Mrate\ vs. $L_{\rm bol}$ relations 
by about 1.5~dex. Considering the 1$\sigma$ scatter reported in \citet{Musiimenta2023}, our object is within 2$\sigma$ from the relation. 
To estimate \Mrate, we assume the electron density of $n_e=1100$~cm$^{-3}$ derived from the \SII\ lines in Sect.~\ref{sec:xshooter_fit} . Assuming  $n_e=200$~cm$^{-3}$ as done by \citet{Fiore2017} and \citet{Musiimenta2023}, our values of \Mrate\ would increase by $\sim0.7$~dex, moving further above the relation.
We note that both relations were derived for objects with $\log L_{\rm bol}/[\text{erg s}^{-1}]>44.5$, thus the relation at low luminosity is just an extrapolation.

\subsection{Emission line ratios}
\label{sec:line_ratios}

Emission-line diagnostics diagrams using NIR line ratios are used to distinguish between different line excitation mechanisms \citep[e.g.,][]{Larkin1998, Oliva2001, Riffel2013,  Colina2015, Maiolino2017, Riffel2021, Calabro2023}.
 In particular, the \FeII$\lambda1.257$/\Pab\ flux ratio has been used together with the \Htwo$\lambda2.12$\micron/\Brg\ flux ratio to separate different excitation mechanisms such as star formation, AGN, and shocks \citep{Larkin1998, Riffel2013, Riffel2021}.\footnote{We note that we cannot apply the new NIR diagnostics from \citet{Calabro2023} as they involve the \SIII$\lambda 9530$\AA\ line which is not in the ERIS spectral range for our target.} Sources in this line diagnostic diagram typically lie along a diagonal line, where low line ratios correspond to star formation, intermediate line ratios suggest AGN activity, and high line ratios are associated with shocks or LINERs activity.

Given the limited number of lines covered by the ERIS data, we focus on the \FeII/\Pab\ flux line ratio, whose map is shown in Fig.~\ref{fig:map_ratio_FeII_Pab}. Additionally, we show a map colour-coded by the classification categories based on \citet{Bianchin2024}, which adapted the \citet{Riffel2013} classification: star formation ($\log \text{\FeII/\Pab}<-0.22$), AGN with low ($-0.22<\log \text{\FeII/\Pab}<0$) and high excitation ($0<\log \text{\FeII/\Pab}<0.3$), and shocks ($\log \text{\FeII/\Pab} >0.3$). We note that these classification diagrams  also include the \Htwo$\lambda2.12$~\micron/\Brg\ ratio, whereas our classification is based solely on the \FeII/\Pab\ ratio, since the \Htwo\ and \Brg\ lines are not covered in the ERIS J\_low band spectral range. Given the known correlation between these two line ratios \citep{Riffel2013}, the \FeII/\Pab\ ratio alone still provides valuable diagnostic information.

The \FeII/\Pab\ map of \MCG\ shows a minimum in the central part of the galaxy, between the north and central \OIII\ peak, close to the position of the continuum peak (see Fig.~\ref{fig:map_ratio_FeII_Pab}). The line ratios in that region are in the `AGN low-excitation' region of the \citet{Riffel2013} diagram.
The line ratios around the north and south \OIII\ and X-ray peaks are more elevated, placing them in the `AGN high-excitation' area of the diagram; farther out, in the more external regions, the ratios increase further to values consistent with shock excitation.

To investigate the origin of the low \FeII/\Pab\ flux ratio in the central region, 
we extract an integrated spectrum from that region and compare the line profiles of \Pab\ and \FeII\ (see Fig.~\ref{fig:comp_FeII_Pab_profiles}).
 \Pab\ exhibits an excess of emission relative to \FeII\  at low velocities, which is quite narrow. This rules out the presence of a hidden broad line region (BLR) causing the excess \Pab\ emission. 
 The enhancement in the \Pab\ flux is likely driven by star formation, and may indicate the presence of a starburst in the central region. The line ratio does not fall within the star-forming region of the diagnostic diagram, likely due to the overlapping contributions from both star formation and AGN activity.
We note that the region with the lowest ratio corresponds  to the peak of the continuum emission observed with HST and ERIS.
The central starburst could be induced by the merger (given that this galaxy  is an advanced merger stage) or by the outflows compressing the gas toward the central region, producing positive feedback. 
In favour of this scenario, \citet{Miniutti2007} analysed  \textit{XMM-Newton} observations of \MCG\ and found thermal plasma emission in the soft energy band, which they attributed to a recent starburst event of 15-20~\Msunyr.
Another possibility is that the lower ratio is due to the presence of a higher ionising radiation in that position, due to the ionisation cone of an AGN, that increase the \Pab\ flux and not the \FeII. 
Assuming that the ionisation cone follows the same orientation as the  radio jet (NE-SW direction), it would point in the direction of the continuum position.

\subsection{Radio emission}
\label{sec:radio}
In this section, we discuss the radio emission in \MCG\ observed with the VLA at 15~GHz and 33~GHz. 
Figure~\ref{fig:nuclei_position} shows the contours of the VLA 33~GHz emission on top of the \Pab\ flux map. This emission is aligned in the NE-SW direction, extending approximately from the north  to the central \OIII\ peak, and is composed by four main emission spots. These regions were already identified in \citet{Song2022}, who attribute them to AGN and jets, but do not specify the location of the AGN. Indeed, the morphology is consistent with a single small jet extending roughly 180~pc.
The AGN could be located in the brightest northern region.
Alternatively, the AGN location could be in the southern part, with the  jet pointing towards the NE, and the counter-jet not visible. Or it could be located in the central peak, whereas the extended aligned emission corresponds to the two jets.

To identify the region from which the jet originates, we calculate the spectral index ($\alpha$) from the 15~GHz and 33~GHz data, defined as $\alpha=\frac{\log(F_{15\text{GHz}}/F_{33\text{GHz}})}{\log (15\text{GHz}/33\text{GHz})}$.
We first smoothed the 33~GHz map to match the resolution of the 15~GHz image ($0.24\arcsec \times0.11\arcsec$) using the CASA task \textsc{imsmooth}. We applid a small astrometric shift to align the peak of the two images ($\Delta$Dec.$ =0.035\arcsec$).
We produced a spectral index map, shown in Fig.~\ref{fig:radio_images}. 
We also estimated the spectral index in two circular regions  with radius 0.15\arcsec (larger than the beam size), centred on the north and south part of the jet (see Fig.~\ref{fig:radio_images}). The spectral index is flatter in the south part ($\alpha=-0.7\pm0.1$) and steepens in the north region ($\alpha=-1.2\pm0.1$).
The radio cores have usually flat spectral indices  ($\alpha=(-0.6,1.6)$, $<\alpha >=0.22\pm0.03$), while the jets can have steeper indices \citep[mean $<\alpha>=-1.04\pm0.03$,][]{Hovatta2014}.
The steepening rate of this jet is consistent with other radio jets \citep[][]{Hovatta2014}.
The spectral index map suggests that the jet most likely originates in the southern region. However, the radio emission peaks in the north, contrary to the expectation that the core should coincide with the radio flux maximum. Alternatively, the peak of the radio flux in the north could be due to a hotspot. 
As a third possibility, if the radio AGN is in the central spot, the difference in the spectral indices of the two regions can be explained by the fact that they are two separate jets.
Despite the fact that the location of the origin of the radio jet remains unclear,  the morphology of the radio emission suggests the presence of a single radio AGN. Thus, the second AGN, if present, would  be radio quiet.

\subsection{X-ray spectra}
\label{sec:Xray_spectra}
\begin{figure}[!t]
\centering 
\includegraphics[width=0.45\textwidth]{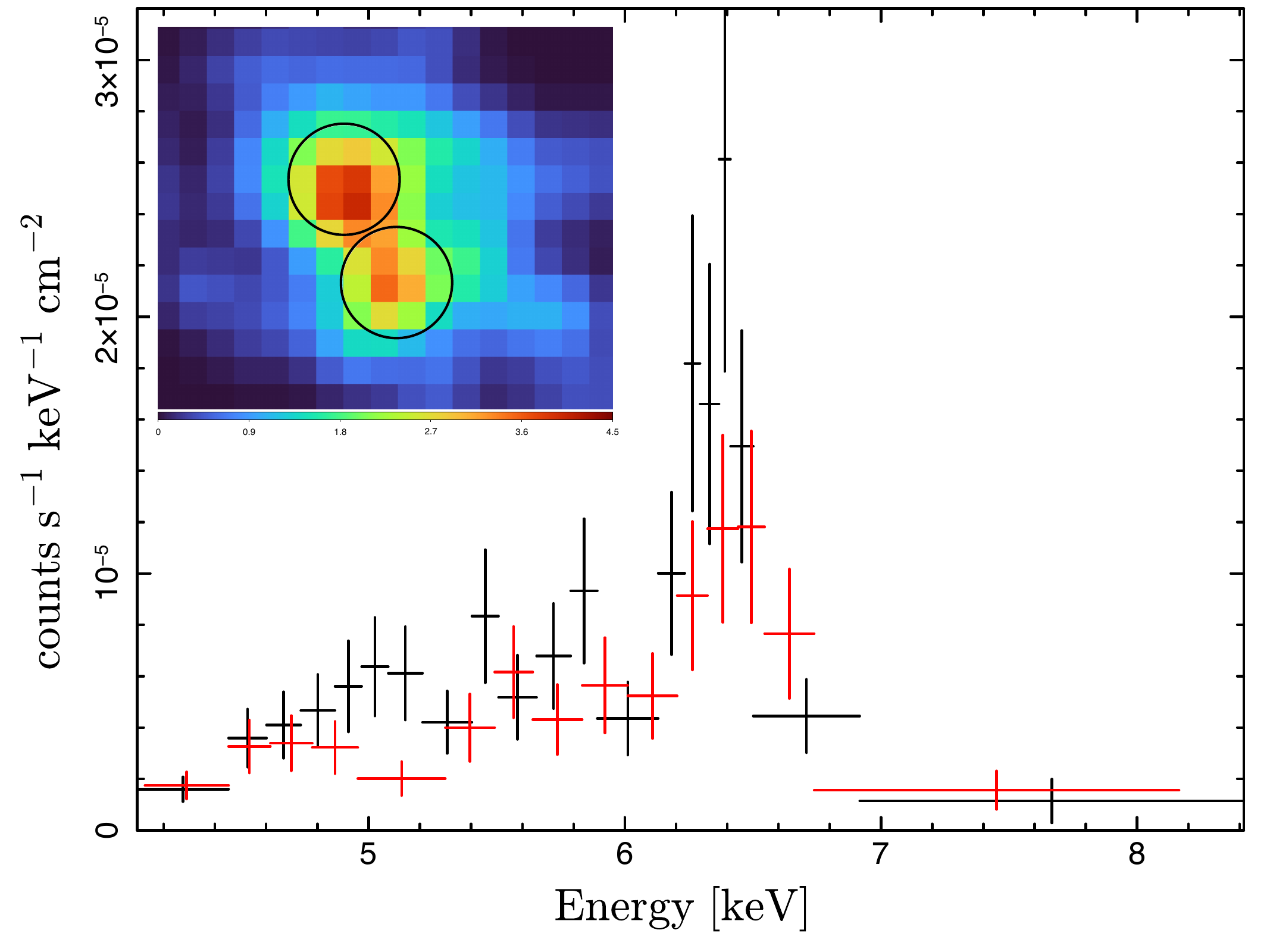}
\caption{
Hard X-ray emission of the north (black) and south (red) Fe-K$\alpha$ clumps, binned to a 3$\sigma$ significance per spectral channel for clarity. Both spectra are suggestive of a heavily obscured AGN. The inset panel shows the \Chandra\ image in the Fe-K$\alpha$ band (same as in Fig. \ref{fig:nuclei_position}), with overlaid the two 0.25\arcsec\ circular regions adopted for a visual inspection of the local spectra.} 
\label{fig:Xray_spectra}
\end{figure}

In order to further investigate the evidence for a dual AGN, we attempt to gain some more pieces of information from the spatially-resolved spectra of the two Fe-K$\alpha$ emitting clumps. We stress that this is a largely qualitative effort, since the extraction of spectra from subpixel regions might not be entirely reliable and some degree of contamination between the two regions cannot be excluded. As anticipated in Sec.~\ref{sec:ancillary_data}, we considered two circular regions with radii of 0.25\arcsec\ that encompass the Fe-K$\alpha$ centroids and the bulk of the emission from the relative clumps (see inset panel in Fig.~\ref{fig:Xray_spectra}). The north (south) region contains 50 (38) counts in the 6.2--6.6 keV rest-frame band, and 515 (364) overall (observed 0.5--7 keV band). A zoom-in on the hard portion of the two resulting spectra is shown in Fig.~\ref{fig:Xray_spectra}. In both cases, the shape is the typical one of heavily obscured AGN, where X-ray photons scattered into the line of sight by a high-column density material dominate over the directly transmitted ones. The north clump appears to be slightly brighter in both the 6.4-keV feature (unsurprisingly) and the lower-energy continuum, which might be therefore the {\it reflection} emission associated with the fluorescent line in both cases. Other subtle differences between the two spectra, although marginally significant on statistical grounds, can be perceived in the soft X-rays, where hints of line emission are present in the north component around the energies of the He-like Mg\,\textsc{xi}, Si\,\textsc{xiii}, and S\,\textsc{xv} triplets, which might be part of the photo-ionised gas component already detected in the system \citep{Miniutti2007}. 
The latter features are not seen in correspondence of the south clump, suggesting that the local environment of the two putative AGN has different physical properties.

The key question then is whether the morphological and spectral properties of the south Fe-K$\alpha$ emitting clump can be accounted for by the X-ray illumination of a single AGN located to the north. By modelling the underlying continuum with a phenomenological curved shape and the Fe K$\alpha$ with a Gaussian profile, we estimate a line-only luminosity of $\approx$\,2.7 and $2.0\times10^{40}$ erg s$^{-1}$ for the north and south clump, respectively.\footnote{These values are slightly lower than those measured by \citet{TrindadeFalcao2024} from the narrow 6.2-6.6~keV band (($3.2\pm0.6)\times10^{40}$ erg s$^{-1}$), as expected since the continuum was not subtracted in their analysis.} As already noted by \citet{TrindadeFalcao2024}, the line-only luminosity of the south clump largely exceeds the entire diffuse hard X-ray luminosity observed in most of the heavily obscured local Seyfert galaxies (e.g., \citealt{Jones2021}).

While X-ray line emission from highly ionised Fe species can arise from shocks associated with fast (several thousand km s$^{-1}$)  outflows (e.g., \citealt{Wang2014}), the 6.4-keV neutral-Fe line is almost exclusively produced via fluorescence. It may be instructive to follow the parallel with the molecular cloud Sgr~B2 in our Galactic centre, whose column density approaches 10$^{24}$~cm$^{-2}$, and which is located at about 130 pc from Sgr~A$^{*}$ (comparable to the separation between the two putative AGN in \MCG). We follow \citealt{Murakami2000} (see their Eq.~4; see also \citealt{Terrier2010}), who estimate the intrinsic luminosity of the primary source required to reproduce the observed cloud flux using an X-ray reflection nebula model. Their model assumes a cylindrically symmetric cloud with a prescribed mass and density profile, illuminated by a primary X-ray source with a power-law spectrum ($\Gamma = 2$). The cloud is discretized into cubic cells, within which absorption, reflection, and \FeKa\ fluorescence are computed using standard cross sections.
Were the south clump a dense molecular cloud, in order to produce its Fe-K$\alpha$ luminosity of $\approx$\,$2\times10^{40}$ erg s$^{-1}$ the intrinsic 2--10 keV luminosity of the north AGN should be of the order of $\approx$\,10$^{45}$ erg s$^{-1}$. 
This is almost an order of magnitude larger than the bolometric luminosity estimated for the north clump by \citet{TrindadeFalcao2024} from the \OIII\ emission, and about two orders of magnitude larger than the corresponding 2-10~keV luminosity. Thus,  assuming that the north AGN was much brighter in the past, and that the properties of the south molecular cloud are extreme compared to those of Sgr~B2, the external illumination scenario is not impossible but requires a high degree of fine tuning.


\subsection{Optical coronal lines: the \FeVII\ line profile}
\label{sec:coronal_lines}

\begin{figure}[!]
\centering 
\includegraphics[width=0.4\textwidth]{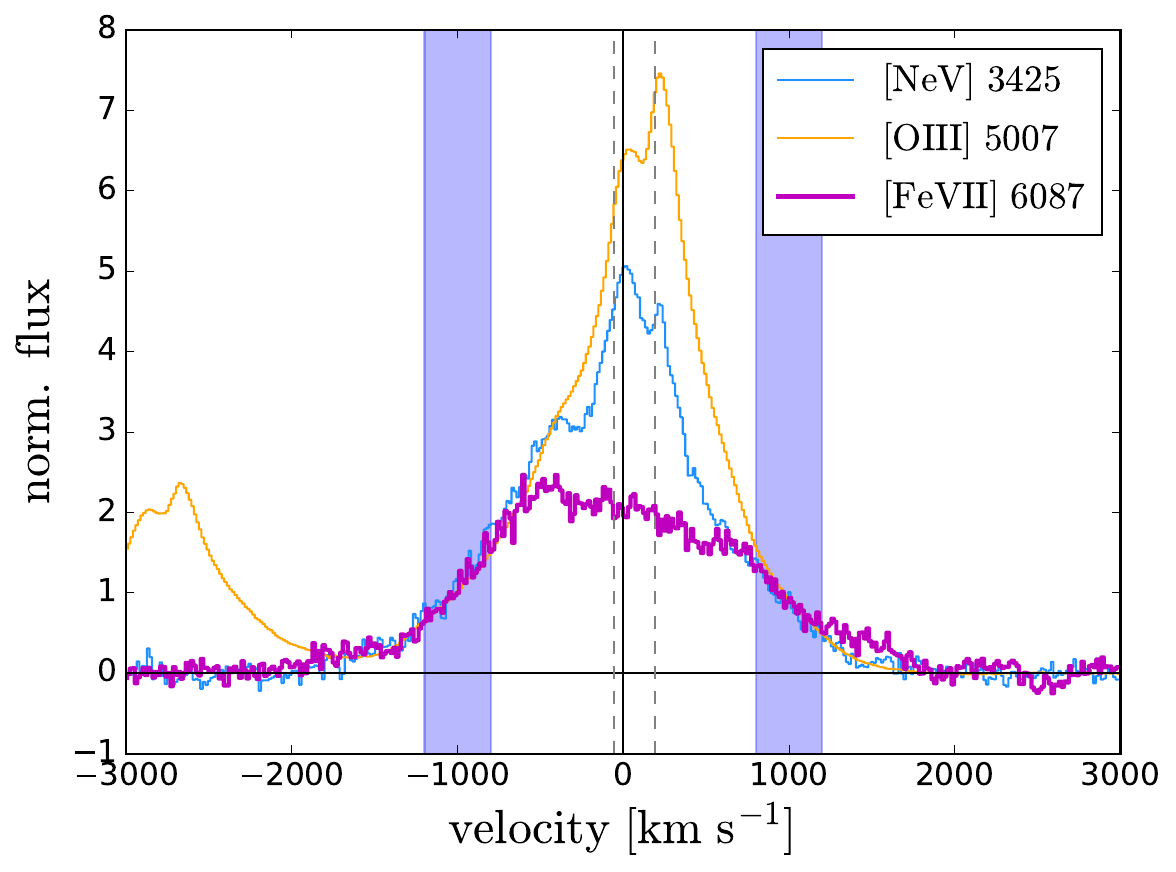}
\caption{Comparison of the \FeVII\ $\lambda6047$ line profile with the \NeV $\lambda3425$ and \OIII $\lambda5007$ line profiles. The lines have been normalised taking the mean flux in the two spectral windows highlighted in violet ($\pm[800, 1200]$~\kms). The black vertical line shows the systemic redshift of the system, the dashed lines show velocities of the two nuclei measured from the ERIS outflow spectra at $-58\pm10$~\kms\ and $190\pm20$~\kms.} 
\label{fig:comp_FeVII_OIII_profiles}
\end{figure}

\begin{figure}[!]
\centering 
\includegraphics[width=0.4\textwidth]{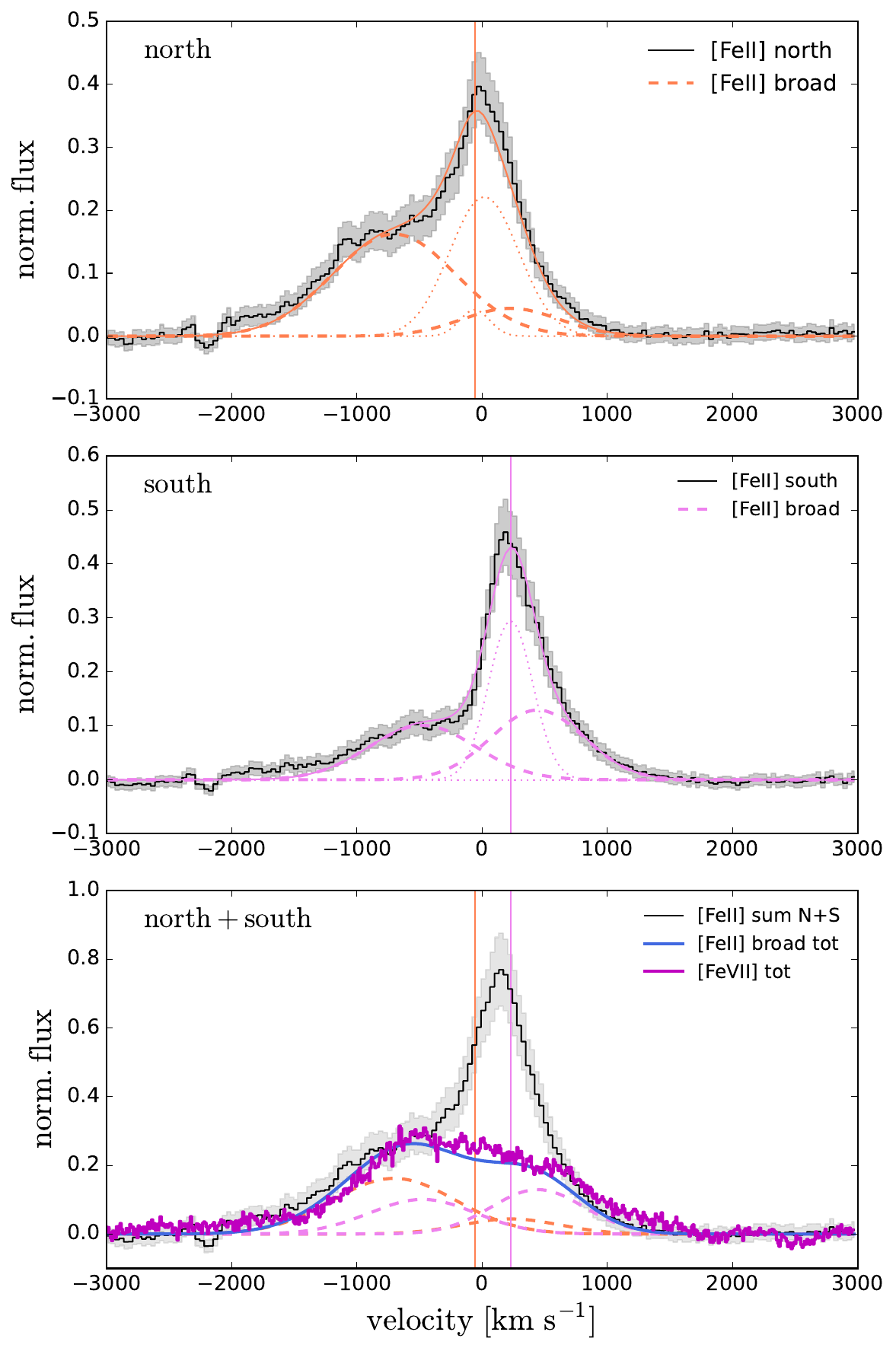}
\caption{Comparison of the \FeVII $\lambda6047$ and the \FeII $\lambda 1.26$~\micron\ line profiles. \textit{Top and middle panels}: \FeII\ spectra extracted from the north and south \OIII\ (aperture radius 0.1\arcsec). The profiles have been modelled with up to 4 Gaussian components (dotted and dashed curves). The `broad' blueshifted and redshifted components (outflow components) are shown with dashed curves. 
\textit{Bottom panel}:
In black is the sum of the north and south \FeII\ spectra.
The blue curve  is the sum of the broad components of the north and south spectra (orange and pink dashed curves), representing the \FeII\ outflow component. 
The \FeVII\ spectrum is shown in magenta, normalised to match the \FeII\ outflow component. This panel highlights the similarity between the \FeVII\ profile and the  \FeII\ total outflow profile.
}
\label{fig:comp_FeVII_FeII_profiles_v2}
\end{figure}

Inspecting the X-shooter optical spectrum of \MCG, we notice that the \FeVII $\lambda6087$\AA\ line has an unusual profile, which is very wide and flat on top.
Figure~\ref{fig:comp_FeVII_OIII_profiles} shows the \FeVII\ line profile compared with the \OIII $\lambda5007$\AA\ and \NeV $\lambda3425$\AA\ line profiles. Line fluxes have been normalised so that the profiles match each other in the high-velocity wings ($\pm[800-1200]$~\kms\ velocity range).
We compare \FeVII\ with \OIII, since it is one of the brightest optical lines, and with \NeV, which has a similar creation ionisation potential to that of \FeVII\ (99~eV for \FeVII\ and 97~eV for \NeV) and a similar critical density  ($1.7\times10^7$~cm$^{-3}$ for \FeVII\ and $1.5\times10^7$~cm$^{-3}$ for \NeV, estimated with {\tt PyNeb}  assuming a temperature of $10^4$~K, \citealt{Luridiana2015}).
Both \OIII\ and \NeV\ show two narrow peaks, that could be related to the emission of the two nuclei, and broad wings reaching high velocities (up to $-1800$~\kms\ in the blue and 1000~\kms\ in the red part).
 The two peaks corresponds roughly to the velocities of the nuclei measured from the narrow components in the ERIS outflow spectra (see dashed lines in Fig.~\ref{fig:comp_FeVII_OIII_profiles}). 
The \FeVII\ matches quite well the profiles of \NeV\ and \OIII\ at high velocities, but does not show the additional narrow emission.
A possible explanation for the peculiar \FeVII\ profile is that this line is produced only in the outflow.

In order to test this hypothesis, we compare the \FeVII\ profile with the \FeII\ outflow emission of the two nuclei. We extract two ERIS spectra from apertures of radius 0.1\arcsec\ centred on the positions of the north and south \OIII\ peaks. We assume that the \FeVII\ peaks at similar positions as \OIII, close to the location of the two nuclei.
We fit the two spectra with up to four Gaussian components: a narrow component ($\sigma<120$~\kms) and three broad ($\sigma>120$~\kms) components (blueshifted, redshifted and systemic). This allows us to isolate the blue- and red-shifted part of the outflow. Then, we sum the outflow components of the two spectra (two blueshifted and two redshifted components) and we compare this \FeII\ `outflow profile' with the \FeVII\ profile (see Fig.~\ref{fig:comp_FeVII_FeII_profiles_v2}). The \FeII\ `outflow profile' is indeed quite similar to the \FeVII\ profile, albeit with a small shift to the blue.

The comparison with \NeV $\lambda3425$~\AA\ is informative since these two lines have a similar ionisation potential 
and critical density.  
The only difference is that iron emission is more sensitive to the presence of dust. In fact, in dusty ISM, iron is depleted, while neon is not  \citep{McKaig2024}.
  This suggests that in the systemic gas traced by the narrow emission there is dust that suppress the \FeVII, while in the outflow there is   dust-poor component which produces \FeVII. 
We note that the $A_V$  measured from the Balmer decrement is lower in the broad blueshifted component ($A_V= 0.3\pm0.5$) than in the narrow component ($A_V= 1.8\pm0.3$), suggesting a lower dust content in the outflow.
 We also note that the excitation is stronger in the outflow than in the systemic component, since in the high ionisation lines, as for example \NeV, the broad component in stronger than the narrow emission. Thus, the absence of  \FeVII\ in the systemic component is likely due to a combination of low excitation and presence of dust. 


This interpretation is consistent with the results by  \citet{Negus2023}, who finds that iron coronal lines (i.e. lines with high ionisation potential $>60-90$~eV) are more common in galaxies with low dust attenuation. \citet{Lamperti2017} and \citet{denBrok2022} investigate the prevalence of coronal lines in nearby  X-ray selected AGN from the BASS sample, and find higher detection rate of coronal lines in Type 1 than in Type 2 X-ray selected AGN, which would also be consistent with a lower presence of coronal lines in more obscured sources.
Moreover, \citet{McKaig2024} use photoionisation models to investigate the role of dust, metallicity,  and ionising spectral energy distribution in the production of optical coronal lines. They find that dust can suppress the emission of most coronal lines by $\sim3$ orders of magnitude, as a consequence of depletion of metals onto dust grains.
Recent studies of large samples of Type 1 quasars from SDSS find a higher prevalence of outflows  in AGN with coronal lines detections \citep{Matzko2025, Doan2025}. Their possible explanation is that the destruction of dust due to the outflow prevents dust depletion and, in turn, enhances coronal line emission. 
They also find that many coronal lines are broader and blueshifted with respect to lower ionisation potential lines, suggesting their emission is more prominent in the outflows. 
All these findings are in agreement with our results.
Under this interpretation, coronal lines from elements prone to dust depletion (e.g., Fe, Ca) are expected to be suppressed in the systemic gas of heavily dust-obscured sources such as U/LIRGs. 


\section{Dual or single AGN}
\label{sec:discussion}
In this section, we review the evidences in favour of the scenario of a dual AGN or a single AGN for our target.
The two  X-ray peaks with similar intensity in the $6.2-6.6$~keV band, tracing the \FeKa\ line, are the strongest evidence for a dual AGN scenario. We note that the {\it Chandra} maps are obtained by binning at subpixel scale (1/4 of of the native {\it Chandra} resolution) to recover the intrinsic sharpness of the mirror's PSF, 
and that the two peaks are not clearly detected (or, in any case, exactly located) in all the individual {\it Chandra} exposures (see Fig.~4 in \citealt{TrindadeFalcao2024}). Deeper {\it Chandra} observations, which have been already approved for the forthcoming cycle 27 (PI: T.\,J.~Turner) and should extend the total available exposure by a factor of 5, are needed to confirm the detection of the two AGN.
 As discussed in Sec.~\ref{sec:Xray_spectra}, if the northern AGN were responsible for the south \FeKa\ peak, its required luminosity would be much higher than the currently observed value.

The two outflows detected in the ERIS data (see Sec.~\ref{sec:outflows}) are consistent with a scenario of a dual AGN. The location of the outflows is roughly coincident with the location of the X-ray peaks.
Alternatively, the double-outflow could be explained by the interaction of the radio jet pushing the gas at the end of the radio lobes and generating two outflows. 
A similar scenario is observed for example in the local AGN NCG 3393 \citep{Finlez2018} and NGC 4151 \citep{May2020}.
In \MCG, both outflows exhibit more prominent blueshifted components than redshifted ones. This asymmetry is difficult to reconcile with a classical bipolar outflow originating from a single AGN. However, the observations could be explained by an outflow inclined near the plane of the sky, producing both blue- and redshifted components, with the redshifted emission hidden by the host galaxy.

The \FeII\ emission, considered to be a tracer of shocks, shows two peaks roughly at the positions of the two outflows (see Fig.~\ref{fig:flux_maps}), which correspond also to the two ends of the radio jet. These shocks could be triggered by the jets or by the outflows themselves.
The \FeII/\Pab\ map (see Fig.~\ref{fig:map_ratio_FeII_Pab}) show lower values in the centre, close to the location of the optical continuum peak and central \OIII\ peak. These lower values could be related to the presence of an AGN, or to a starburst (see Sec.~\ref{sec:line_ratios}).

The radio spectral index map suggests that the AGN responsible for the jet is located in the southern part of the jet, close to the central \OIII\ peak. This is compatible with both the scenario of a single AGN located close to the central \OIII\ peak, and with the scenario of a dual AGN system, in which only one AGN is responsible for the radio emission.
Overall, all the features that we report in these paper can be explained by both the scenario of a single AGN or a dual AGN, except for the two \FeKa\ peaks.

\section{Summary and conclusions}
In this paper we presented VLT/ERIS IFS observations of \MCG, a candidate dual AGN at redshift  $z=0.0167$ with nuclear separation of $\sim100$~pc.
From the analysis of the ERIS data, we identify two fast outflows, showing  peaks in the $W80$ maps close to the position of the two X-ray peaks.
The  \FeII\ morphology also shows two  clear peaks, close to the position of the north and south \OIII\ centroids.
 We also analyse ancillary optical, radio and X-ray data.  
The main results of this work are:
\label{sec:conclusions}
   \begin{enumerate}

      \item We identified two blueshifted outflows related to the two X-ray positions and the north and south \OIII\ peaks. 
      From the analysis of the \Pab\ line we derive for the north outflow a velocity $v10=-1210$~\kms, mass $M_{\rm out}=(40\pm8)\times 10^4$~\Msun\ and mass outflow rate \Mrate$=20\pm5$~\Msunyr, while for the south outflow we find $v10=-1140$~\kms, $M_{\rm out}=(42\pm10)\times 10^4$~\Msun\ and  \Mrate$=21\pm5$~\Msunyr\ (see Sec.~\ref{sec:outflows}). The outflow velocities of the nuclei are comparable to the fastest $2-5$~\% AGN from low-redshift X-ray selected literature samples \citep{Perna2017a,Rojas2020}.
      
      \item The \FeII/\Pab\ flux ratio is lower at the centre, close to the optical/NIR continuum centroid, and higher towards the outskirts.
      According to the diagram by \citet{Riffel2013}, the outskirts are dominated by shocks, while the central region by AGN. The lower \FeII/\Pab\ ratios in the center could be explained by a nuclear starburst boosting the \Pab\ emission or by high ionisation due to an AGN (see Sec.~\ref{sec:line_ratios}).
      
      \item The \FeVII$\lambda6087$\AA\ coronal line shows a peculiar line profile, very broad and flat on top. The most likely explanation is that this line is emitted only in the outflow component which contains less dust, as suggested by the comparison with the \FeII$\lambda1.257$\micron\ and \NeV$\lambda3425$\AA\ line profile  (see Sec.~\ref{sec:coronal_lines}).

   \end{enumerate}

All the multi-wavelength properties of the system, including the two outflows, can be explained by either a single or a dual AGN scenario. The only feature that clearly favours a dual AGN interpretation is the presence of two \FeKa\ X-ray peaks with comparable intensity, although their detection remains tentative with current data. Deeper X-ray observations will be crucial to confirm the presence of the two peaks and to establish the true nature of the system.

\begin{acknowledgements}

We thank the anonymous referee for their constructive feedback.

Based on observations collected at the European Organisation for Astronomical Research in the Southern Hemisphere under ESO programmes P113.26BW.001 and 0102.A-0433.

We acknowledge financial contribution from INAF Large Grant “Dual and binary supermassive black holes in the multi-messenger era: from galaxy mergers to gravitational waves” (Bando Ricerca Fondamentale INAF 2022),  from INAF Large Grant "The Quest for dual and binary massive black holes in the gravitational wave era",(Bando Ricerca Fondamentale INAF 2024) from the INAF project “VLT-MOONS” CRAM 1.05.03.07,  from the INAF Large Grant 2022 “The metal circle: a new sharp view of the baryon cycle up to Cosmic Dawn with
the latest generation IFU facilities”, from PRIN-MUR project “PROMETEUS” financed by the European Union -  Next Generation EU, Mission 4 Component 1 CUP B53D23004750006.
GC and EB acknowledge financial support from from the INAF GO grant 2024 ``A JWST/MIRI MIRACLE: MidIR Activity of Circumnuclear Line Emission''. EB also acknowledges support from the ``Ricerca Fondamentale 2024'' program (mini-grant 1.05.24.07.01). 
MP acknowledges support through the grants PID2021-127718NB-I00, PID2024-159902NA-I00, and RYC2023-044853-I, funded by the Spain Ministry of Science and Innovation/State Agency of Research MCIN/AEI/10.13039/501100011033 and El Fondo Social Europeo Plus FSE+.
GV acknowledges support from European Union’s HE ERC Starting Grant No. 101040227 - WINGS.
ATF was supported by an appointment to the NASA Postdoctoral Program at the NASA Goddard Space Flight Center, administered by Oak Ridge Associated Universities under contract with NASA.
We are grateful to the ESO staff for support during observations and data reduction.

This research has made use of the VizieR catalogue access tool, CDS,
Strasbourg, France.  The original description of the VizieR service was published in A\&AS 143, 23.
This research made use of Astropy \citep{astropy},  {\tt Matplotlib} \citep{Hunter2007}, {\tt NumPy} \citep{VanDerWalt2011}, {\tt SciPy} \citep{SciPy}, {\tt corner} \citep{corner}, {\tt astroquery.vizier}  \citep{Ginsburg2019_astroquery}.   
\end{acknowledgements}

%
%




\bibliographystyle{aa} 
\bibliography{Paper_biblio_full.bib}

\begin{appendix}

\section{Radio images}
\begin{figure*}[!t]
\centering 
\includegraphics[width=0.99\textwidth]{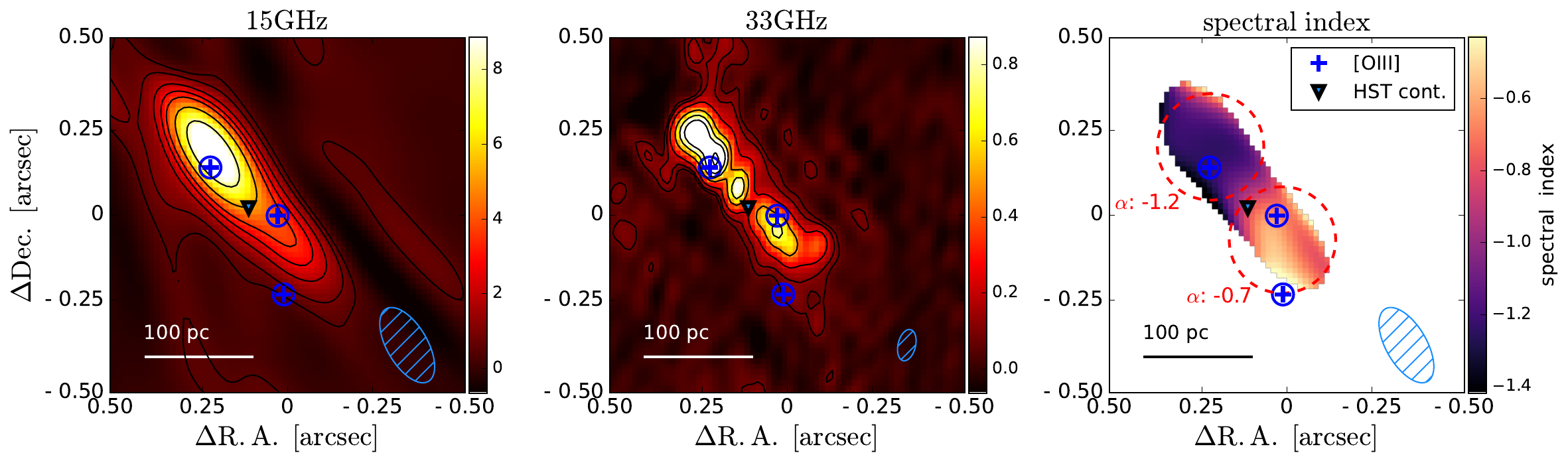}
\caption{Maps of the radio emission from VLA. From left to right: 15~GHz map, 33~GHz map, and spectral index map obtained combining the two radio maps, smoothed at the resolution of the 15~GHz data. The lightblue ellipses show the beam sizes. The symbols show the positions of the multi-wavelength peaks, as in Fig.~\ref{fig:nuclei_position}. }
\label{fig:radio_images}
\end{figure*}

Figure \ref{fig:radio_images} shows the VLA 15GHz and 33GHz continuum maps, and the spectral index map (see Sec.~\ref{sec:radio}). 

\section{ERIS astrometric alignment}
\label{sec:appendix_astrometry}
\begin{figure*}[!t]
\centering 
\includegraphics[width=0.7\textwidth]{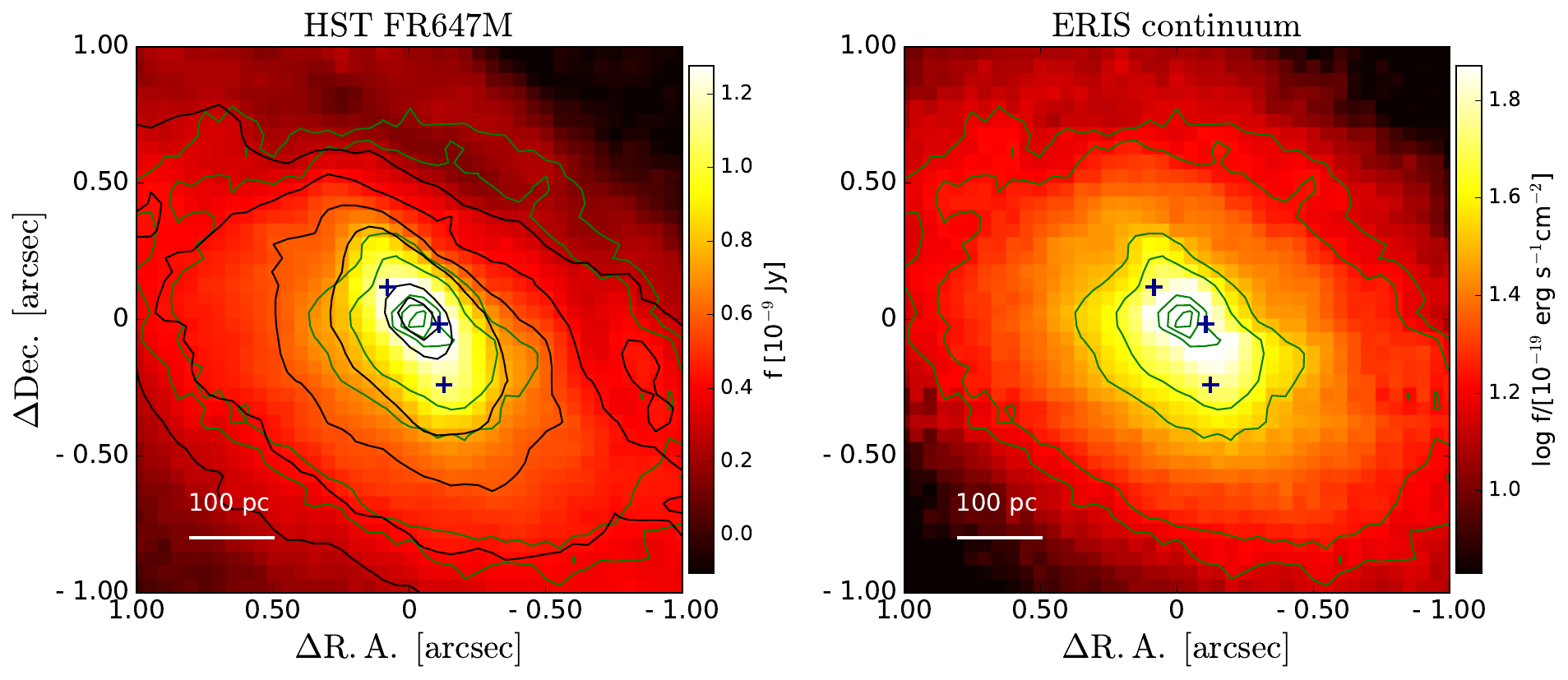}

\caption{Astrometric alignment of ERIS data with HST images. \textit{Left:} HST FR647M continuum image. \textit{Right:} ERIS continuum image, created by collapsing line-free spectral channels. Black and green contours show arbitrary flux levels from the HST and ERIS map, respectively. The three crosses show the position of the \OIII\ peaks.}
\label{fig:HST_ERIS_astrometry}
\end{figure*}

Figure~\ref{fig:HST_ERIS_astrometry} shows the result of the alignment of the ERIS continuum image with the HST F647M continuum image, as described in Sec.~\ref{sec:astrometry}.

\section{Parameters of the multi-Gaussian fit}
In Table~\ref{tab:fit_param}, we report the range and the initial values of the parameters used in the multi-Gaussian fit of the ERIS data-cube.

\begin{table}[h]
\centering
  \caption[]{Range of parameters for the multi-Gaussian fit of the ERIS datacube. All units are [\kms]. }
    \label{tab:fit_param}
    \begin{tabular}{lcccc}
    \hline
        component    &  \multicolumn{2}{c}{velocity} & \multicolumn{2}{c}{$\sigma$}  \\
           &  range & initial & range & initial\\
        \hline
         narrow & [--400, 400] & 0 & [40, 180] & 50\\
         broad 1 & [--850, 700] & --200 & [40, 750] & 300 \\
         broad 2 & [--850, 700] & 100 &[40, 500] & 300\\
    \end{tabular} 
\end{table}

\section{Properties derived from the fit of the X-shooter optical spectrum}
Table~\ref{tab:optical_prop} summarizes the properties derived from the fit of the optical spectrum.
\begin{table}[h]
\centering
\caption{Properties derived from the fit of the X-shooter optical spectrum.}
\setlength{\tabcolsep}{3pt}
\begin{tabular}{lccccccccccc}
\hline
Component & $\delta v$ & $\sigma$ & $A_V$ & $n_e$ &  \\ %
        & [\kms] & [\kms]  & [mag] & [cm$^{-3}$] \\
(1) & (2)  & (3) & (4) & (5)\\ 
  \hline \hline

narrow & $6\pm2$ & $182\pm5$ & $1.78_{-0.25}^{+0.28}$ & $1100^{+200}_{-160}$ \\ 
broad1 & $-400\pm30$ & $700\pm10$ &  $0.28_{-0.28}^{+0.52}$ & -\\
broad2 & $-40\pm10$ & $530\pm20$ & $0.40_{-0.30}^{+0.32}$ & -\\
\hline
broad1+broad2 &  - & - & $0.35_{-0.30}^{+0.68}$ & $950^{+450}_{-250}$ \\ 
total &  - & - & $0.61_{-0.59}^{+0.60}$ &$1000^{+200}_{-150}$ \\ 
\hline

\end{tabular} 
\label{tab:optical_prop}
\tablefoot{
(1) Component of the fit.
(2) Velocity shift with respect to $z=0.01677$.
(3) Velocity dispersion.
(4) Attenuation derived from the Balmer decrement.
(5) Electron density derived from the ratio of the \SII\ doublet using the prescription in \citet{Sanders2016a}. We do not report the $n_e$ values for the individual broad1 and broad2 components, as these two components are strongly degenerate.
}
\end{table}

\section{Electron density}
\label{sec:electron_density}
\begin{figure*}[!t]
\includegraphics[width=0.4\textwidth]
{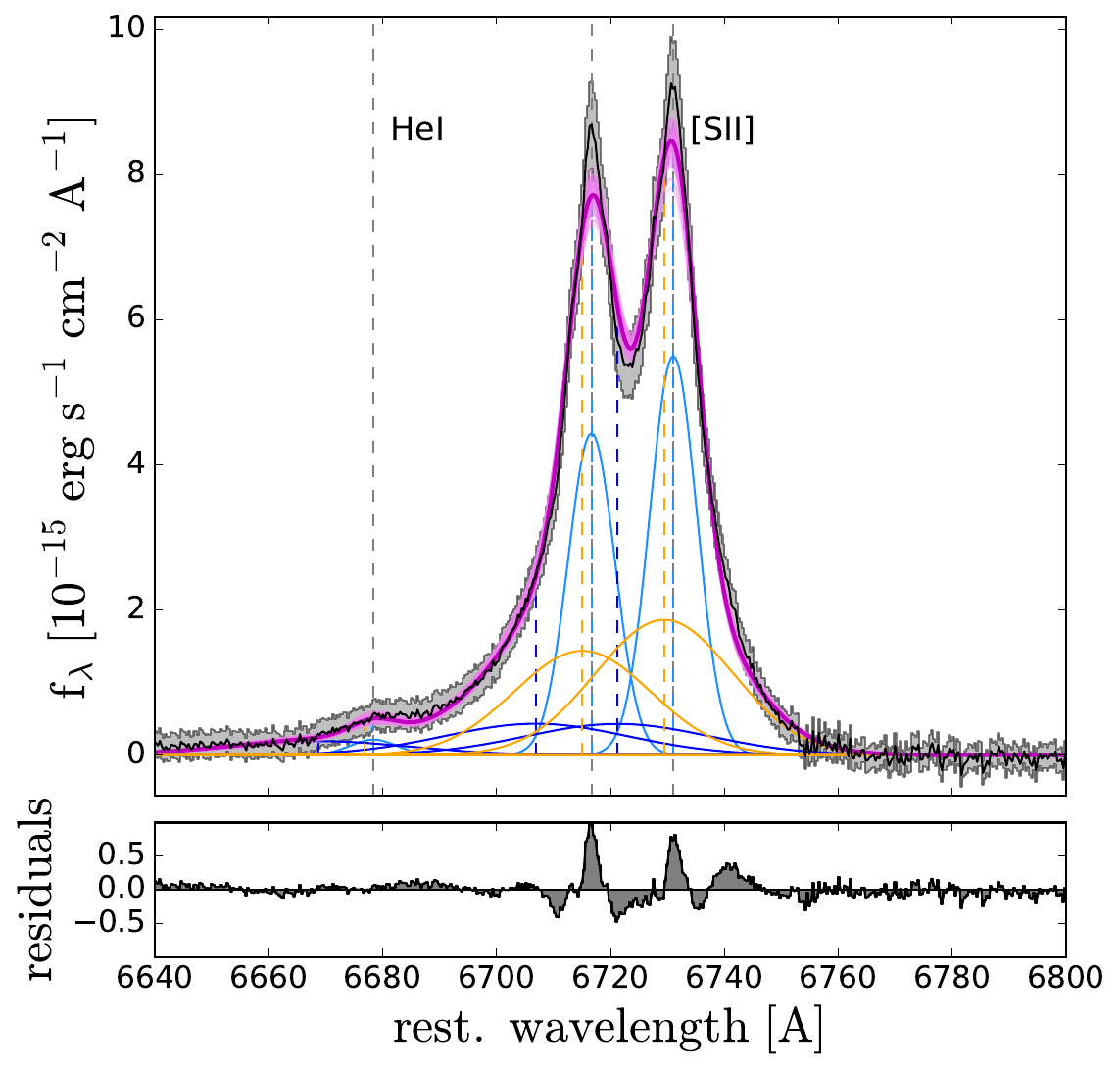}
\includegraphics[width=0.4\textwidth]
{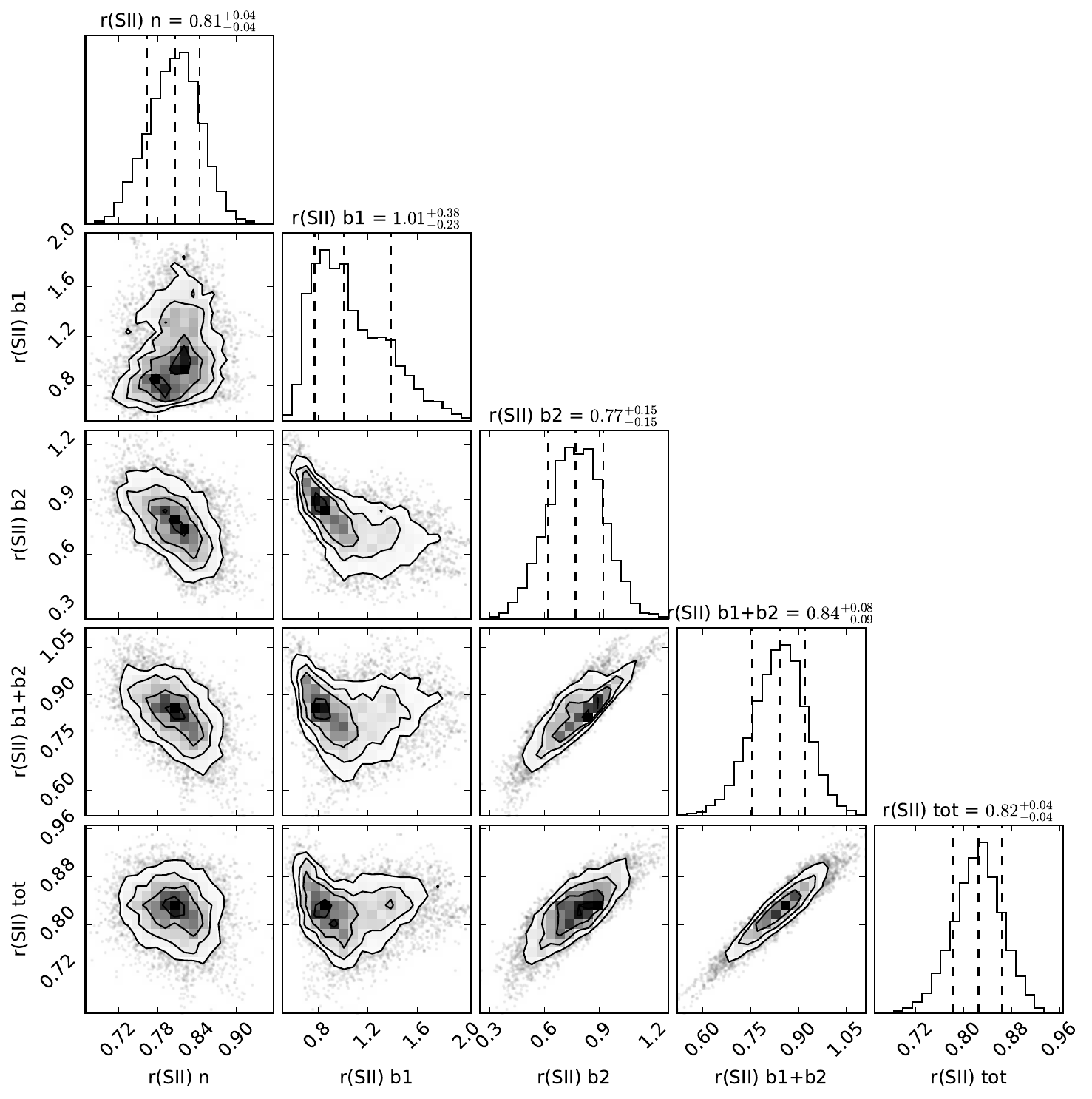}
\caption{Fit of the \SII\ $\lambda\lambda6716,6731$ doublet to derive the electron density. \textit{Left:} Result of the fit of  the \SII\ and \HeI$\lambda$6678  line. The observed spectrum with the corresponding uncertainties is shown in black; the magenta curve shows the best-fit total profile; the light-blue, blue and orange curves show the narrow, broad 1 and broad 2 components, respectively. The bottom panel shows the residuals. 
\textit{Right:} Corner plot showing the posterior distribution of the \SII$\lambda$6716/ \SII $\lambda$6731 line ratio for the different components: narrow (n), broad 1 (b1), broad 2 (b2), combination of the two broad components (b1+b2), and total (n+b1+b2).}
\label{fig:SII_ratio}
\end{figure*}

We estimate the electron density from the \SII$\lambda \lambda$6716,6731 line ratio. In Section~\ref{sec:xshooter_fit}, we describe the fit of the X-shooter optical spectrum using three Gaussian components (narrow, broad 1 and broad 2).
Figure~\ref{fig:SII_ratio} show the best-fit model of the \SII\ doublet and the posterior distribution of the \SII\ ratio parameter. 
We estimate the \SII\ ratio  for the narrow, broad 1 and broad 2 component individually. Additionally, we sum the posterior distribution of the two broad components (broad 1 and broad 2) and of all components (narrow, broad 1, and broad 2), to derive the \SII\ ratio for the total broad profile and the total profile, respectively. These posterior distributions are shown in the corner plot in Figure~\ref{fig:SII_ratio}.


\section{Outflow properties of literature dual AGN}
In Table~\ref{tab:literature_outflows}, we report the ionised outflow velocities and bolometric luminosities of dual AGN from the literature, shown in Fig~\ref{fig:Lbol_vs_vmax}.

\begin{table*}[ht]
\centering
\caption{Summary of dual AGN from the literature with detection of ionised outflows.}
\begin{tabular}{lcccccccccc}
\hline
Name  & $z$ & Type & $d$& $\log L_{\mathrm{bol}}$ & $v_{\mathrm{max}}$ & Ref. & Ref.  \\
 &  &  &  [kpc] & [\ergs] & [\kms] & discovery  &  outflows \\
 (1) & (2) & (3) & (4) & (5) & (6) & (7) & (8) \\
\hline

\MCG\_N & 0.017 & 2 & 0.1 & 44.20 & $1780 \pm 30$ & TF24  & this work\\
\MCG\_S & 0.017 & 2 & 0.1 & 44.00 & $1610 \pm 50$ & TF24  & this work\\

NGC6240\_N & 0.024 & 2 & 0.7 & 44.41 & $674 \pm 96$ & K03  & HM25\\
NGC6240\_S & 0.024 & 2 & 0.7 & 44.90 & $625 \pm 58$ & K03 & HM25 \\
Mrk273\_N$^{*}$ & 0.037 & 2 & 0.74 & 44.11 & $607.6 \pm 1$ & RZ14 & RZ14$^\star$ \\
Mrk273\_SE$^{*}$ &  0.037 & 2 & 0.74 & 43.81 & $833 \pm 8$ & RZ14 & RZ14$^\star$\\
Mrk463\_E &  0.050 & 2 & 3.0 & 45.38 & $773 \pm 30$ & B08 & T18$^{\star\star}$\\
IRAS20210+1121\_S & 0.056 & 2 & 13 & 45.55 & $2160 \pm 380$ & Sa21 & Sa21\\
3C459\_NW$^{*}$ & 0.220 & 2 & 5.3 & 44.82 & $1814 \pm 279$ & B18 & Sp21$^{\star\star}$ \\
GS 10578 &  3.064 & 2 & 4.7 & 45.81 & $1980 \pm 75$ & P25 & V25 \\
LBQS0302--0019 &  3.287 & 1 & 20 & 47.22 & $1209 \pm 9$ & H18 & P23$^\star$ \\
COS1638-A & 3.5057 & 1 & 8.2 & 46.70 & $3551 \pm 131$ & P25 & B25 \\
COS1638-B &  3.5119 & 2 & 8.2 & 46.20 & $2623 \pm 56$ & P25 & B25 \\
COS1656-A & 3.5101 & 2 & 10 & $<45.84$ & $1326 \pm 6$ & P25 & B25 \\
BR1202-0725$\_$QSO &   4.6943 & 1 & 24 & 47.20 & $2400 \pm 90$ & Z25 & Z25\\
BR1202-0725\_SMG &  4.6891 & 2 & 24 & $>44.90$ & $5000 \pm 1000$ & Z25 & Z25\\

\hline

\end{tabular}
\tablefoot{
(1) Name of the nucleus. $^{*}$: Candidate dual AGN according to the authors (see column (9)).
(2) Redshift.
(3) AGN Type.
(4) Projected separation of the two nuclei.
(5) AGN bolometric luminosity.
(6) Outflow velocity calculated as $v_{max}= |v_{broad}-v_{narrow}|+2\times \sigma_{broad}$,  from a fit with two Gaussian components, following \citep{Fiore2017}.
All velocities are estimated from the \OIII$\lambda$5007\AA\ line, except for NCG6240, where \citet{Hermosa-Munoz2025} use the Pf$\alpha$  hydrogen recombination line.
$^{\star}$ For these targets, the authors did not report the values of $v_{max}$ but reported the best-fit parameters, so we re-constructed the line profile based on these parameters,  fitted the profile with two Gaussian and calculated $v_{max}$. 
$^{\star\star}$ For these targets, the authors did not report the $v_{max}$ values or the best-fit parameters, so we extracted the nuclear spectrum from the MUSE data cube, fitted the \OIII\ line, and derived $v_{max}$.
(7) and (8) References for the discovery of the dual AGN and for the paper presenting the outflow detection: 
TF24 \citet{TrindadeFalcao2024},
K03 \citet{Komossa2003},
HM25 \citet{Hermosa-Munoz2025},
RZ14 \citet{RodriguezZaurin2014},
B08 \citet{Bianchi2008_Mrk463},
T18 \citet{Treister2018},
Sa21 \citet{Saturni2021},
B18 \citet{Balmaverde2018},
Sp21 \citet{Speranza2021},
V25 \citet{Venturi2026arXiv},
H18 \citet{Husemann2018a},
P23 \citet{Perna2023},
P25 \citet{Perna2025},
B25 \citet{Bertola2025},
Z25 \citet{Zamora2025}.
}
\label{tab:literature_outflows}
\end{table*}

\section{Separation of kinematic components}
\label{sec:appendix_kin_maps}

\begin{figure*}[!]
\centering 
\includegraphics[width=0.9\textwidth]{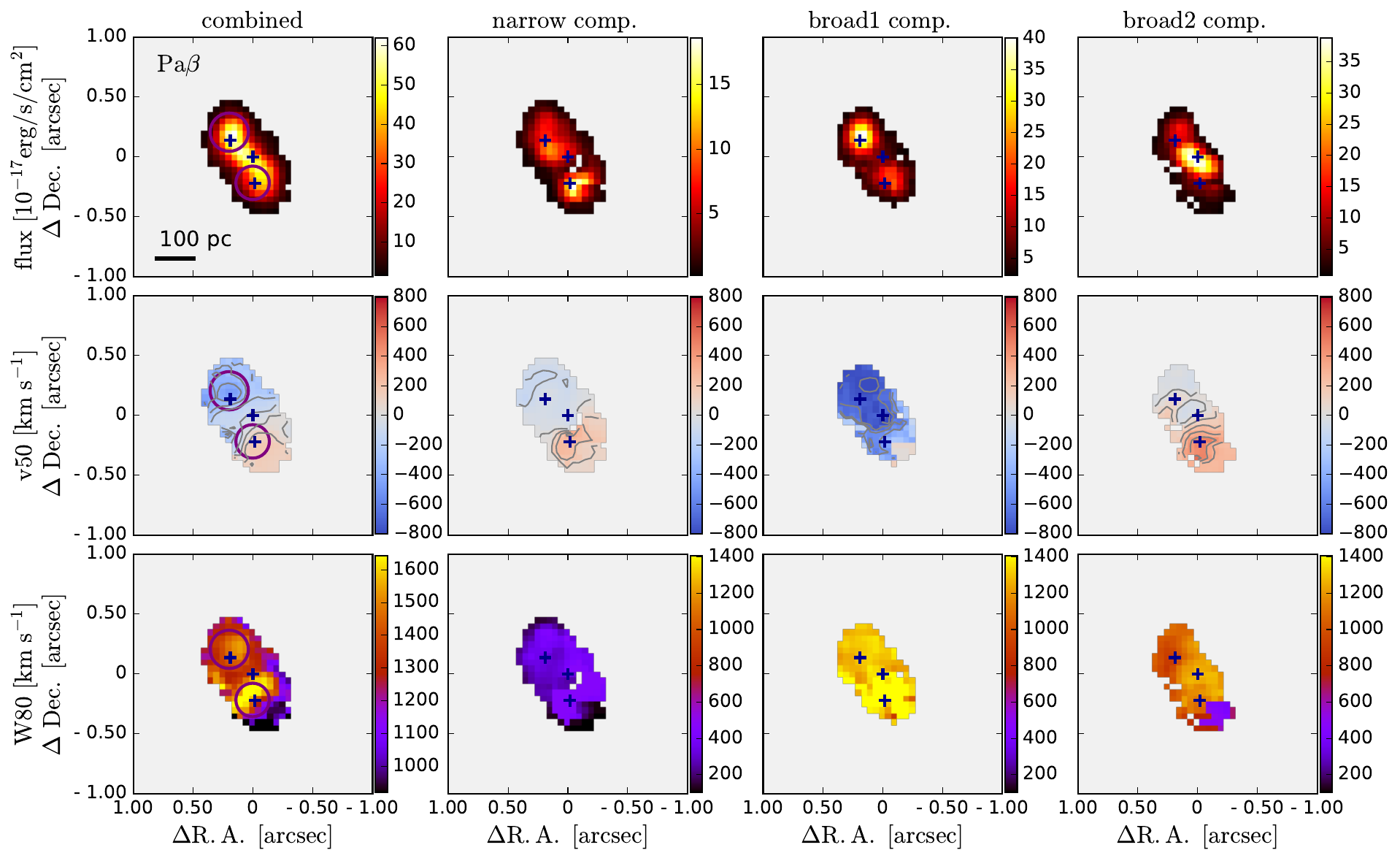}
\caption{Results of the fit of \Pab\ with three Gaussian components. From left  to right: total profile, narrow component, first broad component (b1), second broad component (b2). The top panels show the flux, the middle panels the velocity and the bottom panels the line width W80. 
In the kinematic maps, only spaxels with S/N>3 in the total profile  are shown.}
\label{fig:kin_map_Pab}
\end{figure*}

\begin{figure*}[!t]
\centering 
\includegraphics[width=0.9\textwidth]{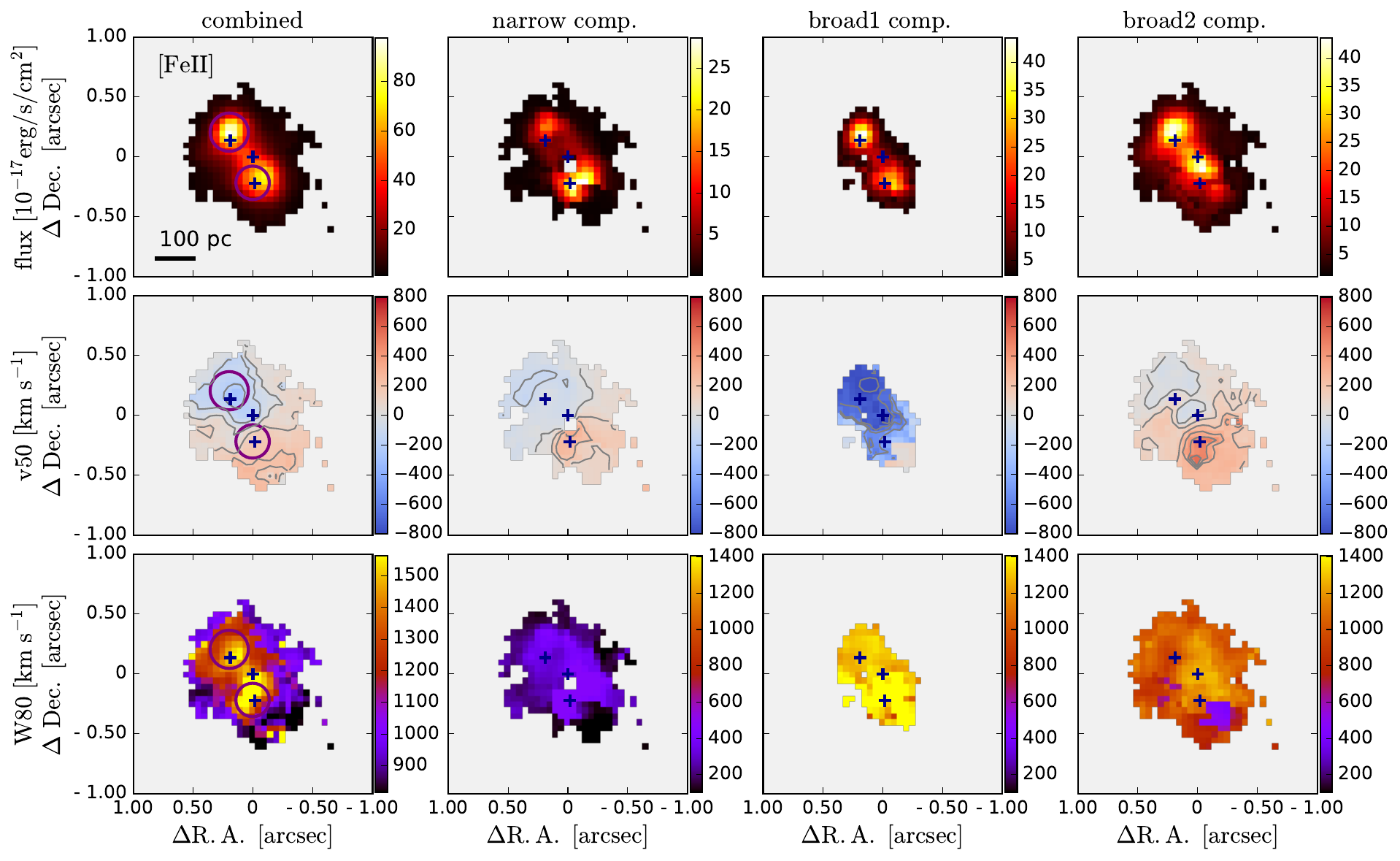}
\caption{Results of the fit of \FeII\ with three Gaussian components. See caption of Fig.~\ref{fig:kin_map_Pab}.}
\label{fig:kin_map_FeII}
\end{figure*}

\begin{figure*}[!]
\centering 
\includegraphics[width=0.9\textwidth]{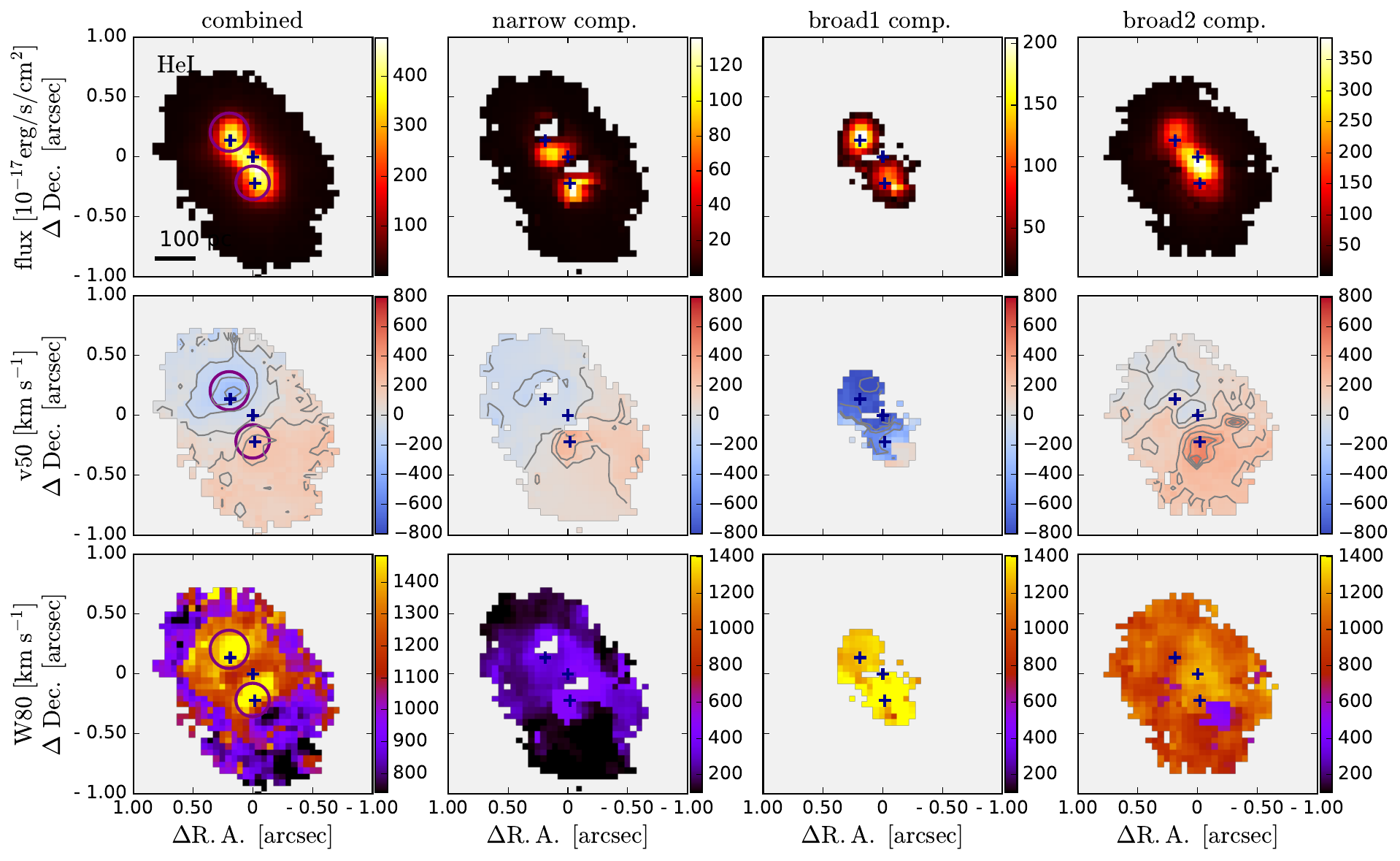}
\caption{Results of the fit of \HeI\ with three Gaussian components. See caption of Fig.~\ref{fig:kin_map_Pab}.}
\label{fig:kin_map_HeI}
\end{figure*}

To reproduce the emission line profiles, we use up to three kinematic components, as explained in Sec.~\ref{sec:analysis}.
In Fig.~\ref{fig:kin_map_Pab}, \ref{fig:kin_map_FeII}, and \ref{fig:kin_map_HeI}, we show the flux, velocity and $W80$ (equivalent to $1.1\times$FWHM for a Gaussian profile) 
maps of the three components (narrow, broad1, broad2), together with the maps obtained from the total profile (sum of the three components) for \Pab, \FeII$\lambda1.257$\micron\ and \HeI, respectively.



\label{fig:FeII_Pab_ratio_map_components}

\end{appendix}
\end{document}